%% file: mkj.tex
\documentclass[useAMS,usenatbib,usedcolumn]{mn2e}


\input ifpdf.sty

\ifpdf
  \usepackage[pdftex,colorlinks,urlcolor=blue,draft]{hyperref}
\else
  \usepackage[draft]{hyperref}
\fi

\usepackage{graphics}
\usepackage{relsize}
\usepackage{amsbsy} 
\usepackage{amssymb}
\usepackage{amsmath}
\usepackage{colortbl}

\newcommand{\unit}[1]{\,\mathrm{#1}}

\newcommand{\oAnd}{{\scriptscriptstyle \rm M31}}
\newcommand{\oMW}{{\scriptscriptstyle \rm MW}}

\newcolumntype{d}[0]{D{.}{.}{1}}

\newlength{\wminus}
\newcommand{\sgcc}[1]{\multicolumn{1}{>{\columncolor[gray]{0.9}}c}{#1}}

\def\hou{^{\rm h}}
\def\min{^{\rm m}}

\newcommand{\vect}{\boldsymbol}
\newcommand{\matr}{\mathbfss}
\newcommand{\transpose}{\mbox{${}^{\text{T}}$}}

\begin{document}
\title[The spatial distribution of MW and M31 satellite galaxies (acc. for pub.)]{The spatial distribution of the Milky Way and Andromeda satellite galaxies}

\author[M. Metz, P. Kroupa and H. Jerjen]
{Manuel Metz$^{1}$\thanks{E-mail: \url{mmetz@astro.uni-bonn.de}}, Pavel Kroupa$^{1}$, Helmut Jerjen$^{2}$\\
$^1$Argelander-Institut f\"ur Astronomie\thanks{Founded by merging of the
\emph{Sternwarte}, \emph{Radioastronomisches Institut}, and \emph{Institut f\"ur Astrophysik
und Extraterrestrische Forschung der Universit\"at Bonn}}, Universit\"at
Bonn, Auf dem H\"ugel 71, D--53121 Bonn, Germany\\
$^2$Research School of Astronomy and Astrophysics, ANU, Mt. Stromlo Observatory, Weston ACT 2611, Australia
}

\date{Received 23 June 2006 / Accepted 17 October 2006}
\pagerange{\pageref{firstpage}--\pageref{lastpage}} \pubyear{2006}

\maketitle
\label{firstpage}

\begin{abstract}
There are two fundamentally different physical origins of faint satellite galaxies: cosmological sub-structures that contain shining baryons and the fragmentation of gas-rich tidal arms thrown out from interacting galaxies during hierarchical structure formation. The latter tidal-dwarf galaxies (TDG) may form populations with correlated orbital angular momenta about their host galaxies. The existence of TDGs is a stringent necessity because they arise as a result of fundamental physical principals. We determine the significance of the apparent disc-like distribution of Milky Way (MW) satellite galaxies. The distribution of the MW satellites is found to be inconsistent with an isotropic or prolate DM sub-structure distribution at a 99.5~per cent level including the recently discovered UMa and CVn dwarf spheroidal galaxies. The distribution is extremely oblate and inclined by about $88\degr$ with respect to the the MW disc. We also apply the methods to Andromeda's (M31) satellite galaxies using two recently published data-sets. It can not be excluded that the whole population of M31 companions is drawn randomly from an isotropic parent distribution. However, two subsamples of Andromeda satellites are identified which have disc-like features. A kinematically motivated subsample of eight Andromeda satellites forms a pronounced disc-like distribution in both data-sets. The existence of this disc would be inconsistent with a CDM parent distribution of subhaloes if the disc is rotationally supported. The M31 satellite distribution is inclined by about $59\degr$ with respect to the M31 disc, and has virtually the same orientation as the disc derived for the whole M31 satellite sample. We present a new geometric method to set restrictions on possible locations of angular momentum vectors for Andromeda satellites. Our conclusion is that both, the MW and M31, may indeed have satellite galaxies derived from TDGs. Further, both host-discs and both identified discs-of-satellites are highly inclined relative to the supergalactic plane. The discs-of-satellites therefore cannot be created from individual accretion events from the supergalactic plane further supporting the possibility that they are of TDG origin.
\end{abstract}


\section{Introduction}\label{sec_intro}
It is well known that in cold-dark-matter (CDM) simulations of large-scale structure in the Universe the growth of structure proceeds via a sequence of hierarchical collapses driven by gravity.  Instabilities, perturbations, and torques in the accretion process generate filamentary-like networks that agree well with observed distributions of galaxies from large-scale redshift surveys. Systems coalesce onto these filamentary networks, which in turn merge to form the structure we see today.

There are, however, several issues on smaller scales which are difficult to address using CDM simulations. The number of observed satellite galaxies around the Milky Way (MW) and Andromeda (M31) is significantly smaller than the number predicted by models \citep[][but see also \citealt{kase06}]{moore99, klypi99, gover04}. This so-called `sub-structure crisis' is usually addressed by invoking small-scale baryonic processes or baryonic-dark-matter biases \citep[and references therein]{kazan04a}, but even then the central density profiles of the sub-structures remain cuspy, despite tidal heating and destruction in the host halo. 
Even extreme baryon removal cannot evolve a cusped to a cored DM halo \citep{gnedi02}. 
The cuspy profile is in disagreement with the density profiles inferred for well-observed MW dSph satellites that are interpreted to be the most dark-matter dominated objects known \citep{wilki02, kleyn03}, an interpretation that may need revision \citep{munoz05}. Indeed the DM profiles inferred for the dSph satellites by solving the Jeans equation are completely inconsistent with CDM profiles \citep{wilki06,gilmo06}.

The spatial distribution of the Milky Way satellite galaxies has long been known to show asymmetric patterns and probable streams of satellites (e.g., \citealt{kunke76,lynde76,majew94,lynde95,hartw00}, \citealt*{palma02}). \citet*{kroup05} tested the spatial distribution of the satellite system against the null-hypothesis that it is drawn randomly from a spherical distribution of dark-matter dominated subhaloes. They found that this hypothesis can be excluded with very high statistical significance, given that empirical constraints show the MW potential to be spherical \citep[e.g.,][]{fellh06}.

In reply to \citeauthor{kroup05}, \citet{kang05} argued that if the Milky Way satellites follow the distribution of the dark-matter within the MW halo rather than the distribution of substructure selected by present-day mass, then the observed distribution of the MW satellites is consistent with being CDM sub-haloes. They based their argument mainly on the apparent rms-height of the observed disc-like distribution showing that a steeper radial number-density distribution yields a smaller rms-height. \citet{zentn05} used a semi-analytic model to identify luminous satellite galaxies in CDM host-haloes. As a second test they tagged the most massive dark-matter sub-haloes as luminous satellites \citep{stoeh02}. They showed that an isotropic distribution is not the correct null-hypothesis, but that the host haloes are mildly triaxial, tending to be more prolate than oblate. Based on the relative height of the distribution they argued that the MW satellite system is consistent with being CDM substructure, albeit with a low probability.
Similarly, \citet{libes05} identified luminous satellites using a different semi-analytic model for star formation. Using halo merger trees, they found that the distribution of the most massive progenitors is consistent with the observed distribution of the MW satellites in all their simulations and that a spherical parent distribution is not the correct null-hypothesis. In contrast to \citet{zentn05} they also found that the distribution of the most massive sub-haloes at present is significantly different from that of the MW satellites and the most massive progenitors. All of these results offer different solutions to the disc-of-satellites problem, but these simulations are based on CDM models that do not include the dissipative physics of galactic disc-formation. The existence of a disc-galaxy and the orientation of the baryonic disc relative to the satellite distribution need to be postulated.

Similarly to the Milky Way satellite system, the satellites of Andromeda seem to be anisotropically distributed as well \citep*{grebe99,hartw00}. \citet{koch06} addressed this issue by performing an analysis using a great-circle fitting routine. They found a planar-like distribution with low statistical significance. However, for a morphologically motivated subsample, including most dSph/dE satellites, they claim a highly significant polar great plane. \citet{mccon06} showed that the M31 satellite system is significantly skewed in the direction of the MW. They also identified possible ghostly streams \citep{lynde95} of subsamples of satellites based on the intersection of all possible kinematical poles (i.e.\ directions of orbital angular momenta).

Given the problems the CDM hypothesis has in dealing with virtually all aspects of the dwarf-satellite problem, it is useful to step-back and to consider some issues of fundamental physics: dwarf satellite galaxies can have two fundamentally different origins \citep{hunte00}: either they are hierarchical building blocks, DM sub-haloes (in CDM cosmology), containing shining baryons and are not yet merged with a host galaxy \citep[e.g.,][]{read06}, or they are anti-hierarchically formed as tidal-dwarf galaxies (TDGs) in tidal arms thrown out from interacting gas-rich galaxies. While the former dwarfs critically depend on cosmological theory, TDGs are a result of well-established fundamental physical principles, the conservation of energy and angular momentum. TDGs must therefore arise in \emph{any} cosmological theory of structure formation. The formation of TDGs is observed in the local Universe \citep[e.g.,][]{hunsb96,kroup98b,weilb03,walte06}, the TDG candidates having gas masses of up to $10^9 M_{\sun}$ and ongoing star formation. The efficiency of the production of TDGs is expected to be much higher during early cosmological epochs due to the large gas content of the progenitor galaxies. Groups of TDGs originating from one encounter have correlated orbital angular momenta and may therefore later form a disc-of-satellites.

In the following the spatial distribution of both, the MW and M31 satellites, is investigated using the same analysis methods, allowing a direct comparison of the properties of the satellite distributions. Our ansatz is to test the null-hypothesis that the satellites of the MW and M31 are distributed in a disc. Exclusion of this hypothesis would only imply \emph{probable consistency} with the theoretical DM sub-structure distribution, notwithstanding the failure of the CDM hypothesis to account for the inferred profiles and number of the putative DM haloes of the satellites \citep{gilmo06}. A pronounced disc-like distribution would provide strong support for causally-connected satellites, subject to the condition that their angular momenta are correlated.

We describe the mathematical methods used to fit planes (\S\ref{sec_planefit}) and to analyse the data (\S\ref{sec_saosd}). We analyse the spatial distribution of the Milky Way (\S\ref{sec_analysisMW}) and Andromeda (\S\ref{sec_analysisM31}) satellite system, respectively, using two recent data-sets for the latter. A new method based on radial velocity measurements is used to set some constraints on possible kinematic associations of Andromeda satellite galaxies (\S\ref{sec_m31kinematics}). Finally, in \S\ref{sec_bootstrap} we constrain the shape of the parent distributions of the MW and M31, which may be prolate, spherical, or oblate DM haloes, and discuss the results in \S\ref{sec_discussion}.

\section{Techniques}\label{sec_techniques}
\subsection{Plane fitting}\label{sec_planefit}
To fit a plane to the data, an unweighted fitting algorithm, known as an algebraic least-squares (ALS) estimate method \citep[see, e.g.][]{chojn00} or eigenvalue analysis, is incorporated. It is similar to the algorithm used by others \citep[e.g.][]{dubin91,hartw00,libes05}. In this method the centroid of the data points, $\vect{r}_0$, is calculated and an eigenvalue analysis of the moment of inertia tensor $\matr{T}_0$ of the position vectors $\hat{\vect{r}}_i = \vect{r}_i - \vect{r}_0$, $i=1 \ldots n$, $n$ being the number of satellites, relative to the centroid $\vect{r}_0$ is performed. The eigenvector corresponding to the smallest eigenvalue is the normal of the plane and the plane contains the centroid. Incorporating the centroid of the data ensures that we correctly find the plane that has the minimum orthogonal distance to the satellites. Not considering the centroid results in the plane being forced to pass through the coordinate origin, i.e.\ a great circle fit is performed. This is done by most authors when calculating the moment of inertia tensor, but not here. We choose to seek planes without forcing them to go through the coordinate origin (the centre of the host galaxy), because this allows a consistency check: the constituent satellites must orbit within the host potential such that any disc made up of a virialised satellite population must pass near the origin. Test setups showed that the ALS algorithm is sensitive in deriving the correct distance of an artificial plane at an one-sigma level only. We calculate the centroid of the data-points and not the centre of mass of the satellites. Therefore, any found plane which has a distance to the host centre larger than about one disc-height may be interpreted as being unphysical.

Since the ALS method is an unweighted fitting routine, distance uncertainties are accounted for by the applied error (AE) method: all satellites are randomly shifted along their line-of-sight with a normal distribution function as the probability function of the magnitude of the shift. The variance of the normal distribution used for the shifting is derived from the distance uncertainty of the measurements. The random shifting is repeated a large number ($10^4$) of times. For the analysis of the distribution of derived normals of the fitted planes the same analysis techniques as described below (\S\ref{sec_saosd}) for the bootstrap re-sampling method are employed. Note that there are, however, schemes to introduce a weighting in ALS \citep[e.g.,][]{hartw00,zentn05} but as a drawback this also influences the interpretation of the derived axis-ratios.

A second method to fit planes is a weighted fitting routine based on the orthogonal distance regression (ODR) package provided by Netlib (\url{http://www.netlib.org/odrpack}). An unweighted fit with the ODR method provides the same result as the (much faster) ALS method. The weights are distance uncertainties in our application.
This algorithm is a little bit subtle since it also estimates the errors of the fitted parameters. Even though the fitting parameters converge, the error estimate may not, which can be understood as having weak constraints on the fitting parameters. This only happened in special cases and never when applied to the satellites of the Milky Way or Andromeda directly.
Further, the solution provided by ODR is strongly dependent on the weights used. The Cartesian variances $\sigma_x^2$, $\sigma_y^2$, and $\sigma_z^2$ derived from the distance uncertainty $\sigma_r$ are not independent. Using the variances only leads to a different fit than using the full covariance matrix. Thus, we always use the full covariance matrix, accounting for correlations of the components.

One measure of the planarity of the distribution is the flattening-parameter $\Delta/r_{\rm cut}$ as given in \citet{kroup05}. $\Delta$ is the root-mean-square height of the disc and $r_{\rm cut}$ is the furthest distance to the Galactic Centre in the satellite sample. The alternative flattening-parameter $\Delta/r_{\rm med}$ as suggested by \citet{zentn05} is also a measure of the planarity, where $r_{\rm med}$ is the median distance to the Galactic Centre of a sample. The formula to calculate $\Delta/r_{\rm cut}$ for an analytical $r^{-q}, q\in \mathbb{R}$, distribution as given by \citet[their eqn. 1]{kang05} is correct only for $q<3$. The more general formula is given by:
\begin{eqnarray}
\Delta = \left\{
  \begin{array}{l@{\quad:\quad}l}
  \sqrt{ \frac{3-q}{3(5-q)}  \,  \frac{r_{\rm cut}^{5-q}-r_1^{5-q}}{r_{\rm cut}^{3-q}-r_1^{3-q}} } & q\in \mathbb{R},\;q\ne3,5 \; , \\
  \sqrt{ \frac{1}{3(5-q)}  \,  \frac{r_{\rm cut}^{5-q}-r_1^{5-q}}{\ln|r_{\rm cut}|-\ln|r_1|} }     & q=3 \; ,     \\
  \sqrt{ \frac{3-q}{3}  \,  \frac{\ln|r_{\rm cut}|-\ln|r_1|}{r_{\rm cut}^{3-q}-r_1^{3-q}} }     & q=5 \; ,     \\
  \end{array} \right.
\label{eqn_rmsD}
\end{eqnarray}
where $r_1$ is the minimum radius of the distribution. $\lim_{r_1 \to 0} \Delta$ always converges to 0 for $q \geq 3$. \citet{kroup05} derived a \emph{linear} probability distribution $\rho(r) \propto r^{-p}$, $1.8 \le p \le 2.6$ for the Milky Way (see also \citealt{koch06} for Andromeda), such that the spherical \emph{volume} density is $\rho_{\rm sph}(r,\vartheta,\phi) \propto r^{-q}$, $3.8 \le q \le 4.6$. \citet{kang05} argued that the formally measured flattening $\Delta/r_{\rm cut}$ of a plane with infinite number of particles following a power-law distribution decreases with power-law index $q$ and converges to zero for $q \rightarrow 3$. Indeed this \emph{always} converges to zero for $r_1 \rightarrow 0$, $q \geq 3$. $\Delta$ decreases with power-law index $q$, but more importantly, $\Delta$ is strongly dependent on the minimum radius $r_1$. This will also influence \emph{any test} that is based on the measured height alone.
Nevertheless, the argument by \citet{kang05} that a disc-like distribution may be mimicked if the satellite distribution is centrally concentrated with one or two outliers is valid \citep[see also][]{zentn05}, requiring more robust statistical methods to be launched, this being one important aim of this study (\S\ref{sec_saosd} \& \S\ref{sec_bootstrap}).

For the ALS method one can derive the axis-ratios $c/a$ and $b/a$ of the square-roots of the eigenvalues $(\tau_1 \leq \tau_2 \leq \tau_3)$ of the moment of inertia tensor: $c=\sqrt{\tau_1}, b=\sqrt{\tau_2}, a=\sqrt{\tau_3}$. The values $(a,b,c)$ are proportional to the root-mean-square (rms) deviation relative to the eigenvectors of $\matr{T}_0$. In addition we use the ratio $c/b$ which indicates whether the triaxial distribution is more oblate ($c/b<b/a$) or more prolate ($c/b>b/a$). Note that this definition of `triaxial more oblate' and `triaxial more prolate' is different to the definition based on the triaxiality parameter \citep[e.g.][]{franx91}. 
The ratio $c/a$ is a better measurement than $\Delta/r_{\rm cut}$ and $\Delta/r_{\rm med}$ in terms of providing the ratio of the rms length in the direction of the smallest and largest extent, but can only be calculated for the ALS method. However, for a small sample, such as in the dwarf satellite application, quantities like $\Delta$ or the ratio $c/a$ may not be a robust measure because of small number statistics.

\subsection{The distribution of normal vectors of bootstrapped samples}\label{sec_saosd}
In previous works \citep{hartw00,kroup05,kang05,libes05,zentn05,koch06} the planarity of the satellite distribution was quantified based on the `thickness' of the distribution in relation to some measure of the total spatial extent: $\Delta/r_{\rm cut}$, $\Delta/r_{\rm med}$, or $c/a$. This is of course a basic requirement to call a distribution disc-like. However, if we think of a centrally concentrated distribution any small number of outliers will determine the orientation of the plane and we will always end up with a `thin' disc \citep{kang05,zentn05}. Consequently additional information about the robustness of a disc-like distribution is needed to draw some statistically significant conclusions. This can be achieved using a re-sampling technique.

The bootstrapping method allows an estimate of the robustness of a disc-like distribution. The fitting to the re-sampled data is done using the unweighted ALS method. If the satellites are not distributed in a well-defined planar-like sheet, the normals of the fitted planes should show a large scatter. \emph{The amount of scatter of the directions of fitted normals to the bootstrapped samples is the quantity we use to determine the statistical significance of a plane-like distribution}.
To test the robustness of a best fitting plane to a set of data-points we quantify
the spread of distributions of normals obtained from the large number of bootstrapping samples. A well-defined plane or disc of data points (satellites) will lead to a tight clustering of normals on the Galactic sky, while a weak disc will yield normals scattered over a large fraction of the sky.

For detailed analysis of the bootstrapped data we follow methods described in \citet*{fi87}. The methods described below are only valid for a sample of unit vectors representing axial data, i.e.\ undirected data. For the present purpose the orientation of the normal of a plane is arbitrary, that is, the normal vector $\vect{n}$ represents the direction of an axis. A matrix $\matr{M}$ is defined as
\begin{equation}
\matr{M} = 
\left(\begin{array}{ccc}
  \hat{x}_{\rm 1} & \hat{y}_{\rm 1} & \hat{z}_{\rm 1}\\
  \hat{x}_{\rm 2} & \hat{y}_{\rm 2} & \hat{z}_{\rm 2}\\
  \vdots & \vdots & \vdots \\
  \hat{x}_{\rm m} & \hat{y}_{\rm m} & \hat{z}_{\rm m}\\
\end{array}\right)\;,
\label{eqn_M}
\end{equation}
where $(\hat{x}_i,\hat{y}_i,\hat{z}_i)$ are the cartesian components of unit vectors $\hat{\vect{n}}_i$, $m$ is the number of unit vectors. An eigenvalue analysis of the matrix
\begin{equation}
\matr{T} = \matr{M}\transpose \matr{M}
\end{equation}
is performed. The eigenvector corresponding to the largest eigenvalue $\tau_3$ of $\matr{T}$ ($\tau_1 \leq \tau_2 \leq \tau_3$) is the (estimated) \emph{principal axis} $\vect{a}_0$ of the input unit vectors, the normal vectors $\hat{\vect{n}}_i$, and the principal axis corresponds approximately to the mean direction of the normal vectors. In our application, $\vect{a}_0$ is the principal axis of the distribution of normals of fitted planes to the bootstrapped sample and we derive its direction on the Galactic sky. The \emph{shape} parameter $\gamma$ is defined as
\begin{equation}
  \gamma = \frac{ \ln{\left( \tau_3 / \tau_2\right)} }{ \ln{\left( \tau_2 / \tau_1\right)} }\;,
\end{equation}
and the \emph{strength} parameter $\zeta$ is defined as
\begin{equation}
  \zeta = \ln{\left( \tau_3 / \tau_1 \right)}
\end{equation}
\citep{fi87}. These two quantities can be used to characterise a distribution of axial data on a sphere. $\gamma$ describes the `clusteriness' of the distribution. $\gamma=1$ indicates the transition between clustered ($\gamma>1$) and girdled ($\gamma<1$) distributions. $\zeta$ is a continues parameter indicating the strength of concentration: the larger $\zeta$ the more concentrated a distribution is (being clustered or girdled), a uniform spherical distribution has $\zeta = 0$.

In addition the spherical standard distance
\begin{equation}
\Delta_{\rm sph} = 
  \sqrt{ \frac{
  \sum_m \left[ \arccos\left(|\vect{a}_{\rm 0} \cdot \hat{\vect{n}}_i|\right) \right]^2 }{ m }
}\;, \label{eqn_sphstdist}
\end{equation}
is calculated where `$\cdot$' denotes the scalar product of vectors. This is the analogue to the linear root-mean-square distance on the sphere. Since we deal with axial data we have to take the absolute value of the scalar product in Eq.~(\ref{eqn_sphstdist}). The spherical standard distance implies rotational symmetry but can be considered as an estimate of the upper limit of the opening angle of the sample of normal vectors for non rotational-symmetric distributions.

\section{Data analysis}\label{sec_analysis}
For both large spirals of the Local Group, the Milky Way and Andromeda, satellite galaxies within the virial radius of the host galaxy are selected. This distance range is chosen since the satellite system within this radius is assumed to be virialised and the individual satellites are very likely bound to their hosts. Based on Local Group timing arguments \citep{kahn59} one may argue that the dark-matter haloes are more extended, that they even possibly overlap. This makes it difficult to associate some outer dwarf galaxies definitely to one of the L$_*$ spirals or classify them as free floating objects.

\subsection{The Milky Way satellites}\label{sec_analysisMW}
\begin{figure*}
  \resizebox{14cm}{!}{ \includegraphics{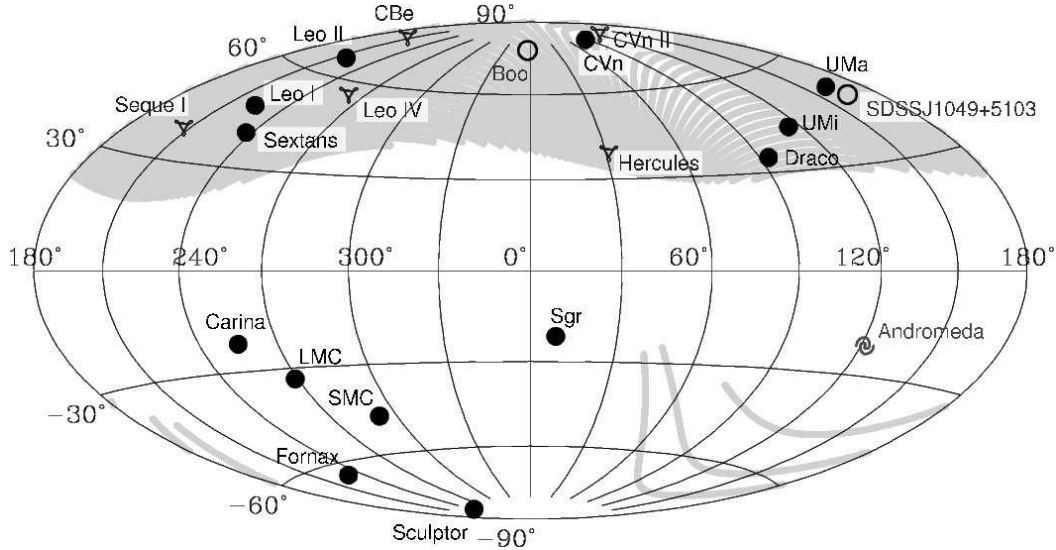} }
  \caption{An Aitoff projection of the positions for the innermost satellites within $254\unit{kpc}$ of the Milky Way as they would appear from the Galactic Centre (compare with Fig.~\ref{fig_satlbm31} for M31). Also the position of Andromeda is shown. The positions of two recently discovered transitional objects are marked by open circles, and the positions of five further very recently discovered companions are marked by triangles. In this projection, the Sun is located at $l_{\rm MW}=180\degr$. The grey shaded region is the full sky-coverage region of the SDSS.}
  \label{fig_satlbmw}
\end{figure*}
For the Milky Way satellite system we use the data-set from \citet[their table~1]{kroup05}, supplementing it with two newly discovered companions: the dwarf galaxies in Ursa major \citep[UMa,][]{willm05} at $\approx 100\unit{kpc}$ and in Canes Venatici at a distance of $220^{+25}_{-16}\unit{kpc}$ \citep[CVn,][]{zucke06}.
Two other recently identified stellar systems, SDSS~J1049+5103 \citep{willm05a} and Bo\"otes \citep[Boo,][]{belok06}, are excluded from the present analysis because of their uncertain physical nature (globular cluster or dSph galaxy). We also exclude the Canis Major dwarf \citep[CMa,][]{newbe02,marti04} at $\approx 15\unit{kpc}$ from the Galactic Centre as its true nature is debated too \citep{moiti06}. The latest discoveries of four further dSph satellites of the MW \citep{belok06a} were not included since the analysis for this paper was finished when the data were published. The positions of the Milky Way satellites are shown in Fig.~\ref{fig_satlbmw} in an Aitoff projection.

\begin{table}
\caption{The positions of the Milky Way satellite galaxies within the approximate virial radius of $254\unit{kpc}$. In the first column we give a running number, in the second the name, in the third and forth longitude and latitude in galactocentric coordinates, and in the fifth the distance with 1-sigma errors from the Galactic Centre. In the sixth column the absolute luminosity in the V-band of the galaxies are given.
}\label{tab_MWdata}
{\centering
\begin{tabular}{cc...@{$\pm$\,}.r}
\hline
No & Name & 
\multicolumn{1}{c}{$l_{\rm MW}$} &
\multicolumn{1}{c}{$b_{\rm MW}$} &
\multicolumn{2}{c}{$r_{\rm MW}$}   & 
\multicolumn{1}{c}{$L_{\rm V}$}\\
 &&
\multicolumn{1}{c}{$[\degr]$} & 
\multicolumn{1}{c}{$[\degr]$} & 
\multicolumn{2}{c}{[kpc]} & 
\multicolumn{1}{c}{$[10^6 L_{\sun}]$} \\
\hline
 1 & Sgr       &   8.8  & -21.5 &   16.0 &  2.0 & $5.1^{\rm(d)}$\\
 2 & LMC       & 269.0  & -33.3 &   50.2 &  2.2 & $2090^{\rm(e)}$\\
 3 & SMC       & 292.2  & -47.1 &   56.9 &  2.2 & $575^{\rm(e)}$\\
 4 & UMi       & 114.5  &  43.1 &   68.1 &  3.0 & $0.29^{\rm(a)}$\\
 5 & Scu       & 237.5  & -82.3 &   79.2 &  4.0 & $2.15^{\rm(a)}$\\
 6 & Dra       &  93.6  &  34.7 &   82.0 &  6.0 & $0.26^{\rm(a)}$\\
 7 & Sex       & 237.1  &  40.5 &   89.2 &  4.0 & $0.50^{\rm(a)}$\\
 8 & Car       & 255.1  & -21.8 &  102.7 &  5.0 & $0.43^{\rm(a)}$\\
 9 & For       & 230.5  & -63.8 &  140.1 &  8.0 & $15.5^{\rm(a)}$\\
10 & LeoII     & 216.5  &  65.5 &  207.7 & 12.0 & $0.58^{\rm(a)}$\\
11 & LeoI      & 223.9  &  48.1 &  254.0 & 30.0 & $4.79^{\rm(a)}$\\
12   & UMa     & 162.0  &  50.8 &  104.9 & 20.0 & $0.04^{\rm(c)}$\\
13   & CVn     &  86.9  &  80.2 &  219.8 & 25.0 & $0.12^{\rm(d)}$\\
\hline
\end{tabular}
}
{\footnotesize References: $^{\rm(a)}$~\citet{mateo98}; $^{\rm(b)}$~\citet{willm05}; $^{\rm(c)}$~\citet{zucke06}; $^{\rm(d)}$~\citet{lee00}}; $^{\rm(e)}$~\citet{vdber99}
\end{table}

\subsubsection{The KTB data-set}
\begin{figure}
  \resizebox{\hsize}{!}{
    \includegraphics{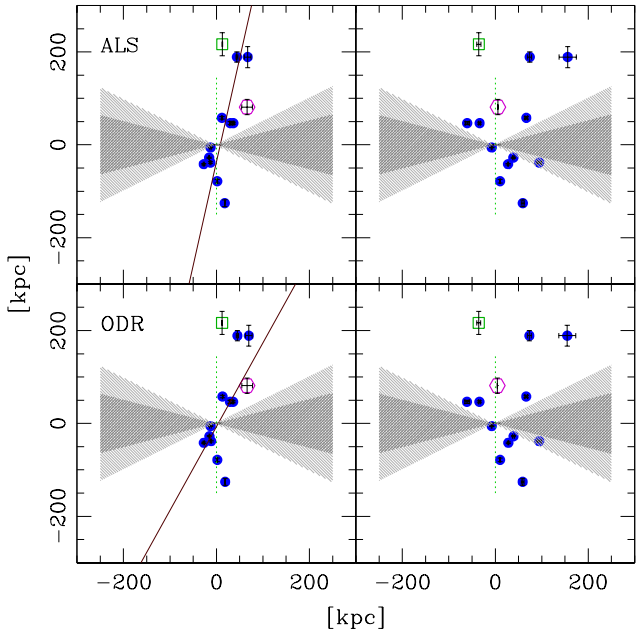}
  }
  \caption{Positions of the MW satellites within the virial radius as seen from infinity. The top panels shows the results of the plane fitting using the unweighted ALS method and the bottom panels for the weighted ODR method when fitting through the complete sample of eleven brightest satellites marked by the filled dots (the KTB sample). In the left panels an edge-on view onto the fitted plane is shown, in the right panel a view rotated by $90\degr$ about the polar-axis of the Galaxy as indicated by the faint vertical dotted line. The MW is located at the origin and its disc is seen edge-on. The position of the UMa dwarf galaxy is marked with an open hexagon, the position of CVn with an open square. The shaded area shows the regions of $b=\pm 15\degr$ and $b=\pm 30\degr$ which may be affected by obscuration through the MW disc.}
  \label{fig_mw11fit}
\end{figure}
First the data-set as used by \citet{kroup05} is analysed. These are the most luminous satellites, comprising a \emph{complete} census of satellites within the virial radius of the Milky Way ($r_{\rm vir,MW}$) and brighter than $M_{\rm tot,V}=-8.8\unit{mag}, L_{\rm V}=2.6 \times 10^5 L_{\sun}$ \citep{mateo98}.
We fit a plane for the innermost eleven satellites (KTB data-set, without UMa and CVn) out to Leo~I with a Galacto-centric distance of $254\unit{kpc}$. For the ALS method, the direction of the normal of the fitted plane is $(l_{\rm MW}=157.3\degr ,\, b_{\rm MW}=-12.7\degr)$ and the distance of the plane from the Galactic Centre is $D_{\rm P}=8.3\unit{kpc}$, which appears to be quite large but is still well within the optical disc of the MW. An edge-on view and a view rotated by $90\degr$ about the polar axis of the Galaxy is shown in Fig.~\ref{fig_mw11fit}. The rms-height is $\Delta=18.5\unit{kpc}$, resulting in flattening $\Delta/r_{\rm cut}=0.07$ and $\Delta/r_{\rm med}=0.23$. So the distance of the fitted plane from the GC is a factor of two smaller than the rms-height of the plane. Note that \citet{kroup05} found $\Delta/r_{\rm cut}=0.10$, and \citet{zentn05} found $\Delta/r_{\rm med} \approx 0.3$ for the same data. The larger values than derived here are caused by the suboptimal fitting routines used there.
The derived axis ratios are $c/a=0.18$ and $b/a=0.53$, resulting in $c/b=0.34<b/a$, a triaxial, highly oblate distribution of satellites. \citet{libes05} gave values of $c/a \approx 0.3$, $b/a \approx 0.5$ for the MW in their figure~3, which would mean $c/b=0.6>b/a$, i.e. a triaxial, slightly prolate distribution.

Next, the applied error (AE) method is applied, randomly shifting the position of all satellites along their line-of-sight vector with a normal distribution function as the probability function of the magnitude of the shift, repeating this $10^4$ times. The principal axis of the resulting distribution of the normals is located at $(l_{\rm MW}=157.4\degr ,\, b_{\rm MW}=-12.6\degr)$ with a spherical standard distance $\Delta_{\rm sph}=1.2\degr$. The derived principal axis is in good agreement with the single fit above and the scatter as quantified by $\Delta_{\rm sph}$ is remarkably small about the principal axis.

Using the ODR method the pole is located at $(l_{\rm MW}=158.2\degr ,\, b_{\rm MW}=-29.0\degr)$ with a distance $D_{\rm P}=3.4\unit{kpc}$ of the plane from the GC. The rms-height is $\Delta=32.6\unit{kpc}$, resulting in flattening $\Delta/r_{\rm cut}=0.13$ and $\Delta/r_{\rm med}=0.41$
The longitude of the derive pole is very similar for the ALS and the ODR method, while the latitude deviate by $\approx 16\degr$. This can be understood by looking edge-on onto the fitted plane (Fig.~\ref{fig_mw11fit}, left panels). Since we are basically sitting in the plane of the satellites, most of the distance uncertainties, which are considered in the ODR method, are along the radius of the plane. As can be seen in Fig.~\ref{fig_mw11fit}, for the distant satellites the components of the distance uncertainties along the polar-axis (the ordinate) are the largest. This forces the ODR algorithm to weight positions along the polar-axis down which results in a different latitude of the pole while the longitude of the fitted normal is not affected. \emph{This outcome is an indication that, while the ODR method is robust against single outliers, it can be biased strongly by a systematic alignment of the provided distance uncertainties.} Later we show that this aspect affects the fitting for the Andromeda satellites even more.

\begin{figure}
  \resizebox{\hsize}{!}{
    \includegraphics{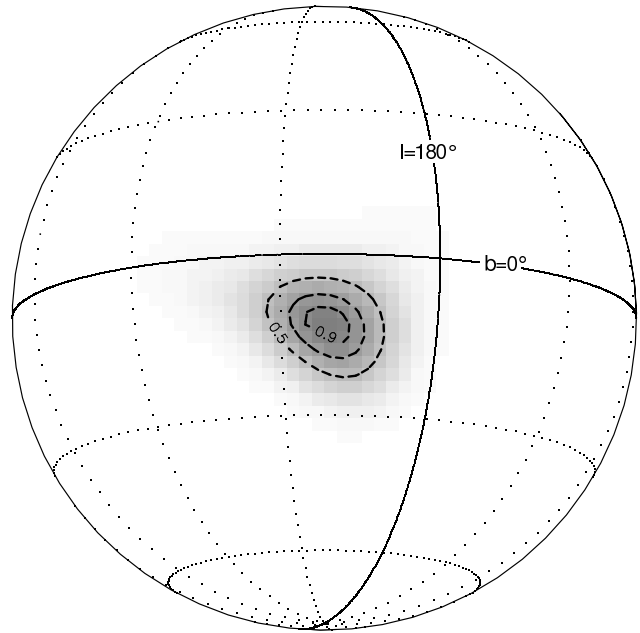}
  }
  \caption{A smoothed ($l_{\rm MW},\,b_{\rm MW}$) scatter plot of the distribution of normals of planes fitted to each of 10\,000 bootstrap samples for the eleven innermost MW satellites (without the recent additions). A spherical density estimate is shown using a Fisher density for the kernel function \citep[$\kappa=100$,][]{fi87} plotted in grey-scales, white corresponds to 0. In addition the contour lines for the density estimate of 0.5, 0.75 and 0.9 in units of the central density are plotted. The principal axis of the distribution is located at $(l_{\rm MW}=158.2\degr,\;b_{\rm MW}=-11.9\degr)$.}
  \label{fig_mw11bootstrap}
\end{figure}

To investigate the robustness of the disc-like feature, 10\,000 bootstrap re-samplings for the eleven innermost satellites of the Milky Way (${^3}N_{\rm tot}=352\,155$ for $n=11$, see Eq.\ \ref{eqn_condbootstrap}) are performed with the ALS method. In Fig.~\ref{fig_mw11bootstrap} a smoothed $(l_{\rm MW},\,b_{\rm MW})$ scatter plot of the locations of the normals of the bootstrap samples is shown in grey-scale, white corresponding to zero density. The smoothing kernel is a Fisher function with smoothing parameter $\kappa=100$ \citep{fi87}\footnote{The Fisher function is the equivalent of the Gaussian on a sphere. The larger the smoothing parameter $\kappa$ is chosen the narrower the smoothing kernel gets.}. In addition the contour lines for the density estimate of 0.5, 0.75 and 0.9 are plotted.
The plot is centred on the principal axis of the distribution which is located at $(l_{\rm MW} = 158.2\degr,\; b_{\rm MW} = -11.9\degr)$ and thus is in good agreement with the results of the single fit above. The shape parameter $\gamma=3.8$ and the strength parameter $\zeta=4.3$ show that the distribution is strongly clustered around its principal axis, and well outside the distribution of $\gamma$, $\zeta$ values for an intrinsically isotropic distribution of satellites (Fig.~\ref{fig_gammazeta}). The resulting spherical standard distance is $\Delta_{\rm sph}=13.0\degr$.

The spherical standard distance $\Delta_{\rm sph}$ found for the AE test is only a tenth of that found with bootstrapping. \emph{This clearly shows that the systematic error caused by the distance uncertainties of the Milky Way satellites is significantly smaller than the intrinsic scatter determined with the bootstrapping.} So the disc-like feature is not affected much by the distance uncertainties of the MW satellites.

\subsubsection{Including the newly discovered satellites}
We repeat the same analysis, now consecutively including the UMa and CVn dwarf galaxies. The results from the single fits, the AE analysis and the bootstrap analysis are given in Tables~\ref{tab_fit}--\ref{tab_resbs}. For the fits the locations of the poles are only marginally affected since both dSphs are located close to the former fitted disc (see Fig.~\ref{fig_mw11fit}). With a distance of $\approx 220\unit{kpc}$ CVn is in fact the second furthest satellite galaxy in our sample. For the unweighted ALS method the orientation of the fitted disc is mostly determined by the outer satellites, nevertheless the orientation is not affected much.

Including both new satellite galaxies, the spherical standard distance $\Delta_{\rm sph}$ found for the AE test remains an order of magnitude smaller than for the bootstrapping, showing that the distance uncertainties do not systematically affect our results.

\begin{table*}
\begin{minipage}{123mm}
\caption{Results from the single plane fits for the innermost satellite galaxies of the Milky Way within the approximate virial radius for the two fitting methods: ALS and ODR (see Text). Results for the KTB data-set, and for the data-sets including the UMa and the CVn dwarf galaxies are given. For Andromeda the results for both data-sets, MI and KG, as well as for a morphologically ({\sf mss8}, \S\ref{sec_mss}) and a kinematically motivated ({\sf kss8}, \S\ref{sec_kss}) subsample of eight satellites are tabulated. Results are given in Galacto-centric and Andromeda-centric coordinates, respectively.}\label{tab_fit}
\centering
\begin{tabular}{lcccccccc}
\hline
method & $l_{\rm MW}\;[\degr]$ & $b_{\rm MW}\;[\degr]$ & $D_{\rm P}\;[{\rm kpc}]$ & $\Delta\;[{\rm kpc}]$ & $\Delta/r_{\rm cut}$ & $\Delta/r_{\rm med}$ & $c/a$ & $b/a$\\
\hline
\multicolumn{9}{c}{\bf Milky Way: KTB}\\
ALS & 157.3 & $-$12.7 &  8.3 & 18.5  &  0.07 & 0.23 & 0.18 & 0.58 \\
ODR & 158.2 & $-$29.0 &  3.4 & 32.6  &  0.13 & 0.41 \\
\multicolumn{9}{c}{\bf Milky Way: KTB + UMa}\\
ALS & 160.5 & $-$14.6 & 12.5 & 20.3  &  0.08 & 0.25 & 0.21 & 0.54 \\
ODR & 158.1 & $-$29.1 &  3.4 & 31.7  &  0.12 & 0.39 \\
\multicolumn{9}{c}{\bf Milky Way: KTB + UMa + CVn}\\
ALS & 153.8 & $-$10.2 &  7.8 & 22.8  &  0.09 & 0.27 & 0.22 & 0.55 \\
ODR & 157.4 & $-$29.2 &  2.9 & 40.9  &  0.16 & 0.48 \\
\hline
 & $l_{\rm M31}\;[\degr]$ & $b_{\rm M31}\;[\degr]$ & $D_{\rm P}\;[{\rm kpc}]$ & $\Delta\;[{\rm kpc}]$ & $\Delta/r_{\rm cut}$ & $\Delta/r_{\rm med}$ & $c/a$ & $b/a$\\
\hline
\multicolumn{9}{c}{\bf Andromeda, MI-data}\\
ALS &  73.4 & $-$31.5 &  1.0 & 45.9  &  0.17 & 0.42 & 0.36 & 0.46\\
ODR &  23.8 & $-$12.5 & 45.0 & 54.4  &  0.20 & 0.50\\
ALS {\sf mss8} &
      177.0 & $-$24.1 & 34.9 & 29.2  &  0.11 & 0.39 & 0.27 & 0.67\\
ODR {\sf mss8} &
      177.9 & $-$20.4 & 37.7 & 29.5  &  0.11 & 0.39 \\
ALS {\sf kss8} & 
       69.9 & $-$35.2 &  1.8 & 16.5  &  0.06 & 0.11 & 0.12 & 0.50\\
ODR {\sf kss8} &
       57.9 & $-$28.6 & 12.8 & 22.7  &  0.08 & 0.16\\
\multicolumn{9}{c}{\bf Andromeda, KG-data}\\
ALS &  83.5 & $-$31.0 &  7.5 & 46.1  &  0.16 & 0.46 & 0.41 & 0.68\\
ODR &  27.6 & $-$31.1 & 32.5 & 68.2  &  0.24 & 0.68 \\
ALS {\sf mss8} &
      168.0 & $-$26.7 &  1.6 &   9.4 &  0.04 & 0.14 & 0.09 & 0.68 \\
ODR {\sf mss8} &
      168.5 & $-$29.4 &  1.9 &  10.2 &  0.04 & 0.15 \\
ALS {\sf kss8} & 
       73.6 & $-$35.0 &  3.9 &  17.9 &  0.06 & 0.18 & 0.15 & 0.71\\
ODR {\sf kss8} &
       52.9 & $-$35.5 & 13.7 &  31.3 &  0.11 & 0.31\\
\hline
\end{tabular}
\end{minipage}
\end{table*}

\begin{table}
\caption{Results from the applied error test for satellite galaxies of the Milky Way and Andromeda. The longitudes and latitudes of the derived principal axis and the spherical standard distance of the distributions for the data-set used by \citeauthor{kroup05}\ are listed. The recently discovered dSph galaxies in Ursa Major and Canes Venatici are included. The results for the two data-sets used for Andromeda, as well as for two subsamples (\S\ref{sec_mss} \& \S\ref{sec_kss}) of M31 satellites are given .}\label{tab_resae}
\centering
\begin{tabular}{c...}
\hline
Data-set & 
  \multicolumn{1}{c}{$l_{\rm MW}\;[\degr]$} &
  \multicolumn{1}{c}{$b_{\rm MW}\;[\degr]$} & 
  \multicolumn{1}{c}{$\Delta_{\rm sph}\;[\degr]$}\\
\hline
Milky Way KTB         & 157.4 & -12.6 &  1.2 \\
                 +UMa & 160.6 & -14.6 &  1.5 \\
                 +CVn & 153.9 & -10.2 & 1.5 \\
\hline
 & \multicolumn{1}{c}{$l_{\rm M31}\;[\degr]$} &
   \multicolumn{1}{c}{$b_{\rm M31}\;[\degr]$} & \\
\hline
Andromeda MI          &  75.1 & -31.7 & 13.2 \\
{\sf mss8}            & 176.6 & -24.9 & 12.5 \\
{\sf kss8}            &  70.5 & -35.2 &  2.4\\
\hline
Andromeda KG          &  82.9 & -31.1 &  8.0 \\
 {\sf mss8}           & 165.2 & -30.9 & 21.5 \\
 {\sf kss8}           &  74.2 & -35.2 & 1.6 \\
\hline
\end{tabular}
\end{table}

\begin{table}
\caption{Results from the bootstrap re-sampling method for the satellite galaxies of the Milky Way and Andromeda. The same quantities as in Table~\ref{tab_resae} are provided and in addition in columns five and six the shape parameter $\gamma$ and the strength parameter $\zeta$ of the distribution of normals of the fitted planes are tabulated.}\label{tab_resbs}
\centering
\begin{tabular}{c.....}
\hline
Data-set & 
 \multicolumn{1}{c}{$l_{\rm MW}\;[\degr]$} & 
 \multicolumn{1}{c}{$b_{\rm MW}\;[\degr]$} & 
 \multicolumn{1}{c}{$\Delta_{\rm sph}\;[\degr]$} & 
 \multicolumn{1}{c}{$\gamma$} & 
 \multicolumn{1}{c}{$\zeta$} \\
\hline

MW KTB   & 158.2 & -11.9 & 13.0 & 3.8 & 4.3 \\
 +UMa           & 161.8 & -14.9 & 12.0 & 4.0 & 4.4 \\
 +CVn           & 156.7 & -10.6 & 12.4 & 2.8 & 4.6 \\
\hline
 & \multicolumn{1}{c}{$l_{\rm M31}\;[\degr]$} &
   \multicolumn{1}{c}{$b_{\rm M31}\;[\degr]$} & \\
\hline
M31 MI    &  75.5 & -31.9 & 38.6 & 0.6 & 3.1 \\
 {\sf mss8}     & 178.3 & -28.5 & 32.7 & 1.1 & 2.7 \\
 {\sf kss8}     &  69.5 & -34.2 &  9.8 & 5.9 & 4.7 \\
\hline
M31 KG    &  83.1 & -30.0 & 27.9 & 2.9 & 2.6 \\
 {\sf mss8}     & 167.1 & -29.6 & 11.8 &  5.3 & 1.9 \\
 {\sf kss8}     &  72.4 & -33.8 & 11.5 & 14.9 & 4.2\\
\hline
\end{tabular}
\end{table}

\subsection{The Andromeda satellites}\label{sec_analysisM31}
In order to study the three-dimensional distribution of the satellite system of Andromeda it is most convenient to transform their position vectors relative to the observer into an Andromeda-centric coordinate system \citep[see also][]{mccon06}. A detailed description of the transformation is given in Appendix \ref{sec_Eccs} in a general way.
Two different data-sets for the distances of Andromeda and its satellites were incorporated: the first data-set as published by \citet[MI data-set, see their table 1]{mccon06} where most of the distances were derived using the tip of the red giant branch method using ground based telescopes \citep{mccon05}. The other data-set as given by \citet[KG data-set, see their table 1]{koch06}: they compiled a list of HST-based distance measurements. The data is given in Table~\ref{tab_M31data} for both data-sets in Andromeda-centric coordinates.

\begin{figure*}
  \resizebox{14cm}{!}{  \includegraphics{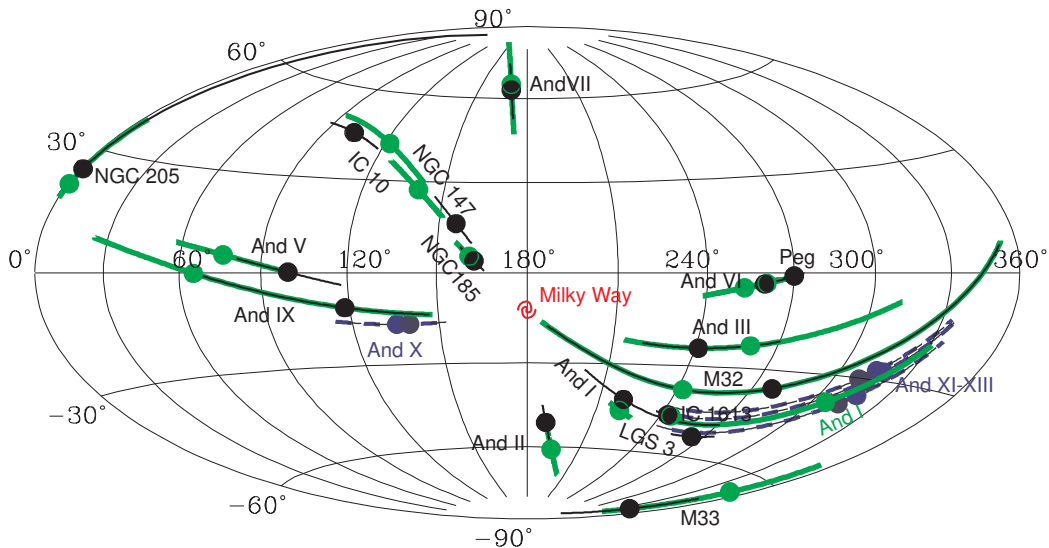} }
  \caption{The Aitoff projection of the M31 satellites in the Andromeda-centric coordinate system for two data-sets: in grey with thick lines for the data given in \citet{koch06} and in black for the data given in \citet{mccon06}. The error bars due to the combined uncertainties in the distance measurement of the satellites and M31 are shown. The position of the MW is also marked.}
  \label{fig_satlbm31}
\end{figure*}

Fig.~\ref{fig_satlbm31} shows an Aitoff projection of the satellite distribution on the Andromeda sky. The error bars due to the combined uncertainties in the distance measurement of the satellites and Andromeda are shown for both data-sets: in grey with thick lines for the KG data-set and in black for the MI data-set. Note the voids in the regions $l_{\rm M31}<180\degr$, $b_{\rm M31}<0\degr$ and $l_{\rm M31}>180\degr$, $b_{\rm M31}>0\degr$. As shown by \citet{mccon06} these regions are only marginally obscured by the MW disc (see their figure 1, the region of maximum obscuration is near IC~10).

Similar to the case of the MW, more faint dwarf galaxies probably remain to be found for M31 within the next few years. One of these discoveries was recently reported by \citet{zucke06b}: the dSph And~X is comparable in luminosity to And~IX, however the distance determination was difficult. \citet{zucke06b} gave a distance of $667 \pm 30\unit{kpc}$ to $738 \pm 35 \unit{kpc}$. Another three satellites, And~XI -- And~XIII, were reported in a very recent paper \citep{marti06}. No distances could be determined for the individual satellites, but combining their colour-magnitude diagrams and assuming all to have the same distance, \citet{marti06} derived a combined distance of 740 -- $955\unit{kpc}$. Given these uncertainties we do not include the four new satellites in our analysis but discuss them later.

\begin{table*}
\begin{minipage}{165mm}
\caption{Basic parameters of the Andromeda satellites: the first four columns contain the a running number, the name, the morphological type and the absolute luminosity in the V-band of the satellite, columns five to seven list the positions in Andromeda-centric coordinates (\S\ref{sec_M31coords}) for the first data-set \citep{mccon06}, and the eighth and ninth columns list the radial and perpendicular components of the measured line-of-sight velocity relative to Andromeda (\S\ref{sec_rpp}). In the columns ten to fourteen the same data are provided for the second data-set \citep{koch06}. The asterisks mark satellites for which the possible poles of the angular momentum vector can be restricted to an arc of $180\degr$ (\S\ref{sec_rpp}).}\label{tab_M31data}
{\centering
\begin{tabular}{ccccddcd@{\;}d@{\,}l%
ddcd@{\;}d@{\,}l}
\hline
 &&&& \multicolumn{6}{c}{MI-data} & \multicolumn{6}{c}{KG-data}\\
No & Name & Type & $L_{\rm V}$ & \multicolumn{1}{c}{$l_{\rm M31}$}
          & \multicolumn{1}{c}{$b_{\rm M31}$}
          & $r_{\rm M31}$
          & v_{\rm t} & \multicolumn{2}{c}{$v_{\rm r}$ }
          & \multicolumn{1}{c}{$l_{\rm M31}$}
          & \multicolumn{1}{c}{$b_{\rm M31}$}
          & $r_{\rm M31}$
          & v_{\rm t} & \multicolumn{2}{c}{$v_{\rm r}$} \\
   &      &      & [$10^6 L_{\sun}$] & [$\degr$] & [$\degr$] & [kpc] 
          & \multicolumn{3}{c}{[${\rm km\,s^{-1}}$]}
          & [$\degr$] & [$\degr$] & [kpc] 
          & \multicolumn{3}{c}{[${\rm km\,s^{-1}}$]} \\
\hline
 1 & M32     & cE & 383$^{\rm(a)}$
              & 278.5 & -35.7 & 6    &  96.0 &   -1.8 & $^{\bigstar}$ 
              & 242.3 & -37.9 & 6    &  83.2 &  -47.9 \\
 2 & NGC 205 & dE & 366$^{\rm(a)}$
              &   0.9 &  25.1 & 40   &  15.0 &   55.1 & 
              &   0.5 &  21.3 & 58   &  11.3 &   55.9 \\
 3 & And IX  & dSph & 0.17 $^{\rm(c)}$
              & 118.3 & -11.1 & 42   &  68.6 &  -51.6 & 
              &  65.4 &  -0.2 & 40   &  82.2 &   24.7 \\
 4 & And I   & dSph & 4.37$^{\rm(b)}$
              & 220.1 & -42.1 & 59   &  71.4 &   38.9 & 
              & 306.2 & -37.8 & 48   &  66.9 &  -46.3 \\
 5 & And III & dSph & 0.58$^{\rm(b)}$
              & 241.1 & -24.1 & 76   &  61.0 &    0.6 & 
              & 260.0 & -22.8 & 68   &  58.4 &  -17.6 \\
 6 & And V   & dSph & 0.58$^{\rm(b)}$
              &  99.6 &   0.2 & 110  & 106.2 &  -38.9 & $^{\bigstar}$ 
              &  76.0 &   5.3 & 117  &  81.0 &  -78.9 \\
 7 & NGC 147 & dE & 131$^{\rm(a)}$
              & 155.8 &  16.1 & 145  &  46.1 & -102.5 & 
              & 122.0 &  42.3 & 101  & 100.3 &  -50.7 & $^{\bigstar}$ \\
 8 & And II  & dSph & 9.12$^{\rm(b)}$
              & 188.7 & -50.9 & 185  &  47.1 & -111.3 & 
              & 193.9 & -60.9 & 160  &  66.4 & -101.0 \\
 9 & NGC 185 & dE & 125$^{\rm(a)}$
              & 162.4 &   3.9 & 190  &  17.5 & -102.1 & 
              & 160.8 &   5.5 & 175  &  21.5 & -101.4 \\
10 & And VII & dE  & 27.5$^{\rm(b)}$
              & 170.1 &  63.3 & 219  &  14.4 &  -84.5 & $^{\bigstar}$ 
              & 169.2 &  65.6 & 216  &  11.0 &  -85.0 \\
11 & IC 10   & dIrr & 160$^{\rm(a)}$
              & 103.2 &  45.2 & 260  &  26.3 & -108.2 & 
              & 140.6 &  27.2 & 255  &  83.3 &  -73.9 \\
12 & LGS 3   & dI/dS & 1.33 $^{\rm(a)}$
              & 264.0 & -53.2 & 269  &   8.8 & -102.1 & 
              & 220.4 & -45.9 & 284  &  41.3 &  -93.8 \\
13 & And VI  & dSph & 3.31$^{\rm(b)}$
              & 259.8 &  -3.5 & 269  &  53.0 & -112.1 & $^{\bigstar}$ 
              & 260.6 &  -3.3 & 266  &  51.4 & -112.8 & $^{\bigstar}$ \\
14 & Peg DIG & dI/dS & 12.0$^{\rm(a)}$
              & 270.1 &  -1.0 & 474  & 145.5 &  -93.0 & 
              & 252.9 &  -4.7 & 410  & 110.7 & -132.6 & $^{\bigstar}$ \\
15 & IC 1613 & dIrr & 63.6$^{\rm(a)}$
              & 243.1 & -46.7 & 511  &  39.2 & -187.3 & 
              & 244.8 & -46.9 & 505  &  43.2 & -186.4 \\
16 & M33     & Sc & 3020$^{\rm(d)}$
              & 340.2 & -77.6 & 207  & 126.3 &  -47.6 & $^{\bigstar}$ 
              & 350.0 & -65.2 & 221  & 133.7 &  -18.5 \\
\hline
\end{tabular}
}
{\footnotesize References: $^{\rm(a)}$~\citet{mateo98}; $^{\rm(b)}$~\citet{mccon06a}}; $^{\rm(c)}$~\citet{zucke04}; $^{\rm(d)}$~\citet{vdber99}
\end{minipage}
\end{table*}

\begin{figure}
  \resizebox{\hsize}{!}{
    \includegraphics{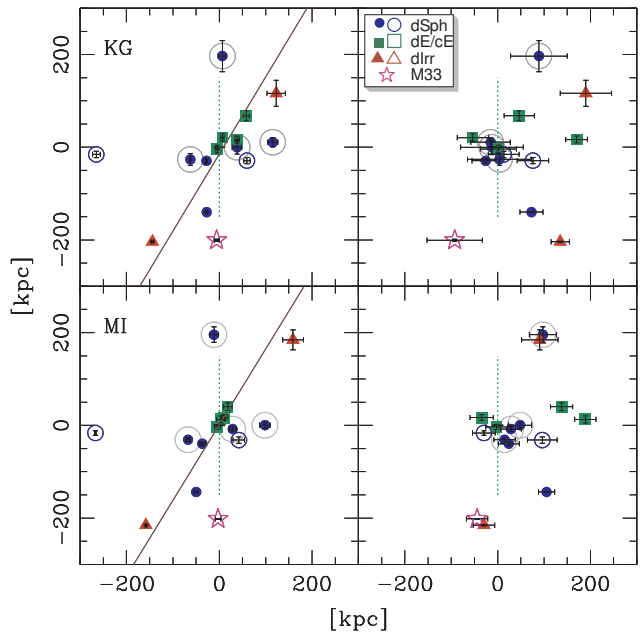}
  }
  \caption{The Andromeda satellite system as seen from infinity. An edge-on view (left panel) and a view rotated by $90\degr$ about the polar axis (right panel) of the fitted planes derived using the ALS method are shown as in Fig.~\ref{fig_mw11fit}. The top panels are the results for the KG-data and the bottom for the MI-data. The different symbols mark different morphological types: dots mark dSphs, squares cE/dEs, and triangles dIrr and dIrr/dSph galaxies. Those satellite galaxies marked with filled symbols were incorporated in the fitting routine while those with open symbols were left out. M33, which was also not incorporated in the fitting, is marked by the pentagram. Satellites encircled by a light grey circle were later excluded from fitting to a kinematically motivated subsample, see \S\ref{sec_kss}. The error bars given are derived from the line-of-sight distance uncertainties of the satellites only.}
  \label{fig_m31fit12}
\end{figure}

The innermost twelve satellites of Andromeda without M33 lie within $\approx 269\unit{kpc}$ which is the approximate virial radius of Andromeda ($r_{\rm vir,M31}$). We assume LGS~3 to be the twelfth satellite, although for the KG data-set And~VI is actually closer to the centre. This will be addressed later (this section and \S\ref{sec_rppapp}). The results are given in Table~\ref{tab_fit}. For the MI-data the distribution is triaxial and more prolate, while for the KG data-set the distribution is found to be triaxial and more oblate, which is reflected in the further analysis, too. Fig.~\ref{fig_m31fit12} shows an edge-on view of the fitted planes and a view rotated by $90\degr$ about the polar axis of Andromeda for both data-sets. As can be seen the recently discovered dSph And~X that was not incorporated in the fitting (marked by the open circle near the centre of the plots) is located close to the fitted plane. If we just recalculate the rms-heights of the fitted disc, now including And~X, it increases only slightly to $\Delta=46.6\unit{kpc}$ for the MI data-set, $\Delta=47.2\unit{kpc}$ for the KG data-set.
Interestingly, also the three very recently discovered satellites And~XI -- And~XIII \citep{marti06} all lie very close ($\approx 10\unit{kpc}$, [$\approx 30\unit{kpc}$]) to the fitted disc when using a mean combined distance of $847.5\unit{kpc}$ albeit with large uncertainties.

For the applied error method (see Table~\ref{tab_resae}) the distance of Andromeda is not varied. This would only affect the distance of the plane from the origin. Compared to the spherical standard distance derived for the MW, it is a factor of ten larger for the M31 system, which is not surprising given the much larger uncertainties in the relative distances M31--satellite. The principal axes come out in good agreement with the poles of the single fits.

However, the ODR method yields poles that are far away from the poles found with the ALS method (Table~\ref{tab_fit}). The difference is totally dominated by the large distance uncertainties, which are here not aligned within the fitted plane as in the case of the Milky Way. Instead they are systematically aligned along the LOS from the MW to M31. The components in the direction to the MW are weighted down and the fitted plane appears to be nearly perpendicular to the direction from the MW to M31, a result of the systematic dependencies of the covariances.

\citet{mccon05} showed that the satellite distribution is significantly offset towards the direction of the Milky Way (also visible in Fig.~\ref{fig_m31fit12}, right panels). The offset is reflected by the large distance from the centre of M31 along the direction of the normal for the fitted planes when using the ODR method due to the nearly face-on orientation of the fitted disc. In contrast, the disc-like distribution found with the ALS method is more edge-on, i.e. the systematic offset as identified by \citeauthor{mccon05}\ is within the plane.

\begin{figure}
  \resizebox{\hsize}{!}{
    \includegraphics{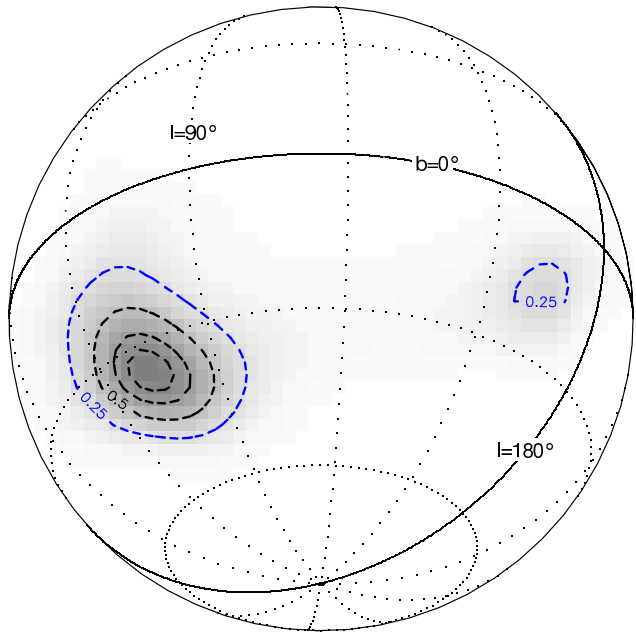}
  }
  \caption{A smoothed ($l_{\rm M31},\, b_{\rm M31}$) scatter plot of 10\,000 bootstrap samples for the innermost twelve Andromeda satellites for the MI data-set ranging out to LGS~3 as Fig.~\ref{fig_mw11bootstrap}. The principal axis of the distribution is located at $(l_{\rm M31}=75.5\degr,\;b_{\rm M31}=-31.9\degr)$. The plot is shown $30\degr$ off-centre from the principal axis. An additional contour line for the density estimate of 0.25 indicates a very weak secondary maximum.}
  \label{fig_m31mccon12bootstrap}
\end{figure}

As for the MW 10\,000 bootstrap re-samplings for both data-sets of M31 are performed to test the robustness of the plane. The principal axis of the distribution is $(l_{\rm M31}=75.5\degr ,\, b_{\rm M31}=-31.9\degr)$ [$(83.1\degr ,\, -30.0\degr)$], being in good agreement with the original fit. We derive a shape parameter $\gamma=0.6$ [$2.9$] and a strength parameter $\zeta=3.1$ [$2.6$], the spherical standard distance is $\Delta_{\rm sph}=38.6\degr$ [$27.9\degr$] (numbers for the MI[KG] data-sets).
While for the KG data-set the distribution of the directions of fitted normals for the bootstrapped sample is found to be clustered ($\gamma > 1$), for the MI-data it is found to be a girdled distribution ($\gamma < 1$). Fig.~\ref{fig_m31mccon12bootstrap} shows a smoothed ($l_{\rm M31} ,\, b_{\rm M31}$) scatter plot of the distribution of the fitted normals for the MI data-set: there is a distinct peak about the principal axis and a second, very weak over-density can be seen nearly $90\degr$ off, being the origin of the girdled distribution. The KG data-set does not show a secondary maximum.

And~VI has approximately the same distance from M31 as LGS~3, both lying close to the approximate virial radius. Including And~VI in the fitting routine for the KG-data, where it is actually closer to the centre of M31 than LGS~3, dramatically changes the picture. The pole of the fitted normals is clearly offset from the fits without And~VI, the distance of the fitted plane is significantly offset from the centre of Andromeda, and the axis ratios do change significantly: $b/a=0.63$ and $c/a=0.57$. More importantly the clustering found for the bootstrapping without And~VI for the KG-data disappears completely. If the satellite was within a planar-like distribution, the bootstrapped distribution should become similarly or even more tightly concentrated as it did when including the UMa dwarf galaxy for the Milky Way. Instead, it gets very weak: $\gamma=0.3$ and $\zeta=1.7$. Therefore we treat And~VI as an outlier.

In contrast to the Milky Way satellite system, for Andromeda the spherical standard distance derived with the applied error method is of the same order as for the bootstrap method which is a result of the large distance uncertainties for M31 and its satellites. So the results may well be affected by the still too large distance uncertainties for M31 and its satellite galaxies.

\subsection{Morphological subsample of Andromeda satellites}\label{sec_mss}
\begin{figure}
  \resizebox{\hsize}{!}{
    \includegraphics{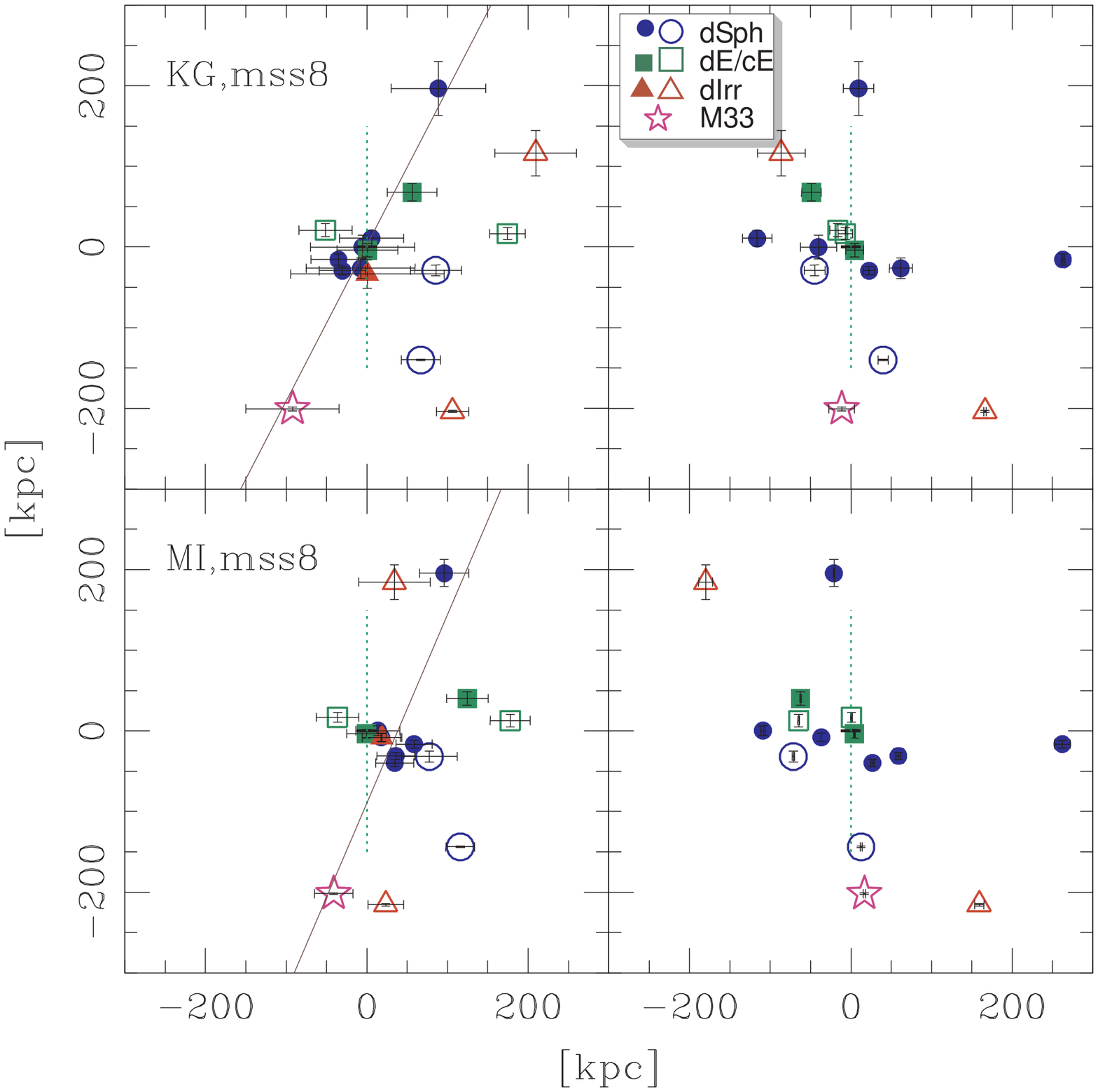}
  }
  \caption{An edge-on view (left panels) and a view rotated by $90\degr$ about the polar axis (right panels) of the fitted plane (using ALS) for the morphologically motivated sub-subsample as proposed by \citeauthor{koch06}. In the top panels we show the results for the MI data-set and in the bottom for KG data-set. Symbols are chosen as in Fig.~\ref{fig_m31fit12}. Satellites not incorporated in the fitting are marked with open symbols. The error bars given are derived from the distance uncertainties of the satellites only.}
  \label{fig_kgsss}
\end{figure}
\citet{koch06} found a very pronounced polar disc-like feature for a morphologically motivated subsample of early-type dwarf galaxies. Their procedure was as follows: they first fitted a plane to all seven dSph galaxies in their data-set and then excluded And~II (at a distance of $r_{\rm M31}=160\unit{kpc}<r_{\rm vir,M31}$) as an outlier because of its large distance to the fitted plane. This disc-like feature of six dSphs was found to be highly statistically significant. Next they included all other morphologically similar galaxies, three dEs and one cE (M32), again finding a disc-like feature with high statistical significance. In a last step they excluded two of the three dEs, but included one transitional type object, the dIrr/dSph Peg~DIG at a distance of $410\unit{kpc}>r_{\rm vir,M31}$, because of its close proximity to the disc-like feature found before\footnote{However, also note the different scaling of the axes in their figure~3 which makes the distributions appear more planar-like than they truly are.}.

Indeed we confirm an amazingly thin configuration when using this sub-subsample consisting of M32, And~I, And~III, NGC~147, And~V, And~VII, And~VI, And~IX, and Peg~DIG. In Tables \ref{tab_fit} -- \ref{tab_resbs} we refer to the sub-subsample without Peg~DIG as {\sf mss8} (morphological subsample of eight satellites). We exclude Peg~DIG because it is well outside the approximate virial radius of M31. The fitted configuration is shown in Fig.~\ref{fig_kgsss}. For the KG data-set the pole of the fitted plane is ($l_{\rm M31}=168.0\degr$, $b_{\rm M31}=-26.7\degr$), with a distance from the centre $D_P=1.6\unit{kpc}$, a rms-height $\Delta=9.4\unit{kpc}$, and with axis ratios $c/a=0.09$ and $b/a=0.68$. For the weighted ODR method the results are very similar (see Table~\ref{tab_fit}). M33 is located very close to this fitted plane.
Including Peg~DIG for completeness results in ($l_{\rm M31}=163.0\degr$, $b_{\rm M31}=-27.3\degr$), with a distance from the centre $D_P=1.2\unit{kpc}$, a rms-height $\Delta=13.1\unit{kpc}$, and with axis ratios $c/a=0.09$ and $b/a=0.45$. 
But this distribution ($b_{\rm M31}=-27.3\degr$) is not as polar aligned as claimed by \citet{koch06} due to the incorrect transformation to the Andromeda-centric coordinate system used by them (\S~\ref{sec_M31coords}). Also note that from the Milky Way we are basically looking face-on onto this fitted plane, which is an important clue as we show later.

Using the MI data-set without Peg~DIG (MI {\sf mss8}) leads to ($l_{\rm M31}=177.0\degr$, $b_{\rm M31}=-24.1\degr$), $D_P=34.9\unit{kpc}$, and $\Delta=29.2\unit{kpc}$; including Peg~DIG, ($l_{\rm M31}=182.4\degr$, $b_{\rm M31}=-23.2\degr$), $D_P=35.1\unit{kpc}$, and $\Delta=29.1\unit{kpc}$. For this data-set the fitted plane is not as thin as for the KG data, and the offset from the centre of M31 is remarkably larger than the rms-height of the fitted plane. As can be seen in Fig.~\ref{fig_kgsss}, lower panel, there is now another dE (NGC~147, the filled square to the right of the plane) remarkably offset from the fitted plane.

Further insight comes from the AE test (Table~\ref{tab_resae}). When applied to the morphological sub-subsample for the derived spherical standard distance is $\Delta_{\rm sph}=21.5\degr$. This is a very large uncertainty in the location of the poles of the fitted normals, a factor three larger than for the full set of twelve satellites within the approximate virial radius used before. For the MI data-set, $\Delta_{\rm sph}=12.5\degr$ is of the same order as for the full data-set.

To test the robustness of the results a bootstrap analysis for the sub-subsample without Peg~DIG is performed, now using 5\,000 re-samplings accounting for the smaller number of possible distinct bootstrap samples (${^3}N_{\rm tot}=6\,231$ for $n=8$). The results are given in Table~\ref{tab_resbs}. The principal axes are in agreement with the single fits. The distribution of direction of bootstrapped normals is marginally clustered and concentrated. However the spherical standard distance for the bootstrapped sample is remarkably smaller than for the AE test. \emph{This indicates that the systematic uncertainties of the distances are larger than the intrinsic scatter of the fitted disc for the KG data.} For the MI data-set the bootstrapped distribution is not found to be clustered but of transitional type. The spherical standard distance is significantly larger than for the KG data-set.

The recently discovered dSph And~X is also found to be off the disc-like structure of the {\sf mss8} subsample for the KG data-set. For a heliocentric distance of $702.5\unit{kpc}$ its distance from the fitted disc is $\approx 87\unit{kpc}$ ($ > 9\sigma$, And~II, excluded as an outlier by \citeauthor{koch06}, is $\approx 127\unit{kpc}$ away). To be within $\pm 10\unit{kpc}$ from the disc, And~X would have to be at a heliocentric distance of $\approx 786$ -- $808\unit{kpc}$. For the MI data-set And~X's distance to the fitted disc is $\approx 32.7\unit{kpc}$. And~X is located on the near side of M31 to the MW, thus adding to the systematic offset of the M31 satellite system towards the barycentre of the Local Group \citep{mccon06}.\\

We thus find that the apparent disc-like configuration of the dSph/dE satellites sub-subsample for M31 is present for the KG data-set only and can not be reproduced using the MI data-set. The nearly face-on alignment relative to the Milky Way results in distance uncertainties basically perpendicular to the fitted plane (Fig.~\ref{fig_kgsss}). The thin configuration disappears when shifting the satellites along their line-of-sight in accord with the distance uncertainties as done in the AE test. Comparing the results for the AE test and the bootstrapping suggests that the systematic uncertainties caused by the distance-measurement errors are larger than the intrinsic scatter of the distribution. Thus the thin disc-like configuration found may be just a chance alignment for the KG data-set, but its existence is not completely ruled out.

\section{Kinematics of the M31 satellites}\label{sec_m31kinematics}
\subsection{Restricted polar paths}\label{sec_rpp}
\begin{figure}
  \resizebox{\hsize}{!}{
    \includegraphics{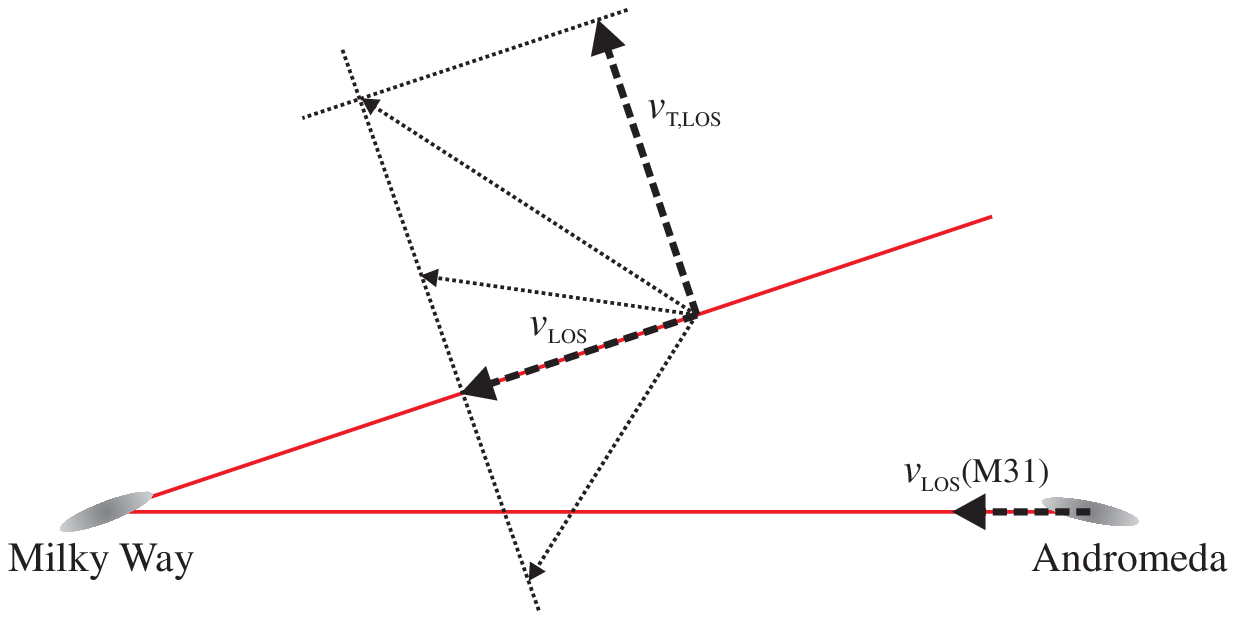}
  }
  \resizebox{\hsize}{!}{
    \includegraphics{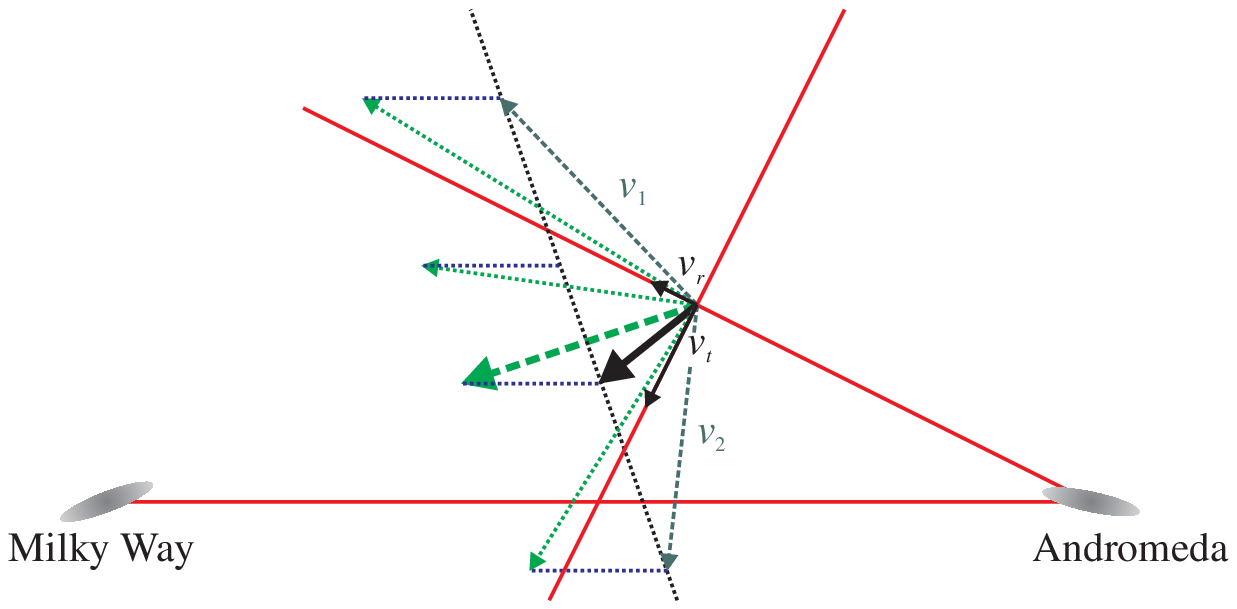}
  }
  \caption{Geometrical illustration how to calculate radial and perpendicular velocity components of the measured line-of-sight (LOS) velocity of an Andromeda satellite. In the top panel we show the measured vectors in the rest-frame of the Sun. The velocity perpendicular to the line-of-sight, $\vect{v}_{\rm T,LOS}$, is unknown, i.e. there is a family of possible total velocity vectors, indicated by dotted arrows. In the bottom panel we show the velocities corrected for the LOS velocity of Andromeda. The corrected LOS velocity of a satellite can be split into a radial component $\vect{v}_r$ and a perpendicular component $\vect{v}_t$ relative to M31.}
  \label{fig_m31vel}
\end{figure}

Based on geometrical arguments, \citet{lynde95} constructed a diagram of so-called `polar paths' for the Milky Way satellite galaxies and globular clusters. The direction of every possible pole of a satellite orbit must be located at a right angle to its direction from the Galactic Centre. Regions in the sky where three or more polar paths (nearly) intersect define the poles of possible streams of satellites. \citet{palma02} used proper motion measurements to restrict the possible orbits, which resulted in arc segments of possible poles rather than great circles. \citet{mccon06} used the available three-dimensional spatial data for the M31 satellites to construct polar paths for Andromeda and identified multiple possible stream candidates of satellite galaxies. We note that a \emph{a disc-of-satellites is rotationally supported if the angular momenta of its constituent satellites are aligned}. Since in a non-spherical halo the angular momentum vectors would precess, the mutual distances of the directions of the angular momentum vectors will grow in the course of time.

The observer's view of the Andromeda satellite system is very different from that of the Milky Way. While the MW satellites are basically seen from the Galactic Centre and only the radial velocity component is available, the M31 satellite system is almost seen from infinity. Together with the spatial information relative to M31 it is therefore possible to decompose the observed line-of-sight (LOS) velocities into radial and perpendicular components relative to M31. These are, however, only lower limits of their total velocities since only the LOS velocities can be measured.

Fig.~\ref{fig_m31vel} illustrates how the measured line-of-sight velocity of an M31 satellite breaks up into radial and perpendicular components relative to M31. The top panel shows the measured heliocentric velocity vectors $\vect{v}_{\rm LOS}$ and $\vect{v}_{\rm LOS}(\rm M31)$.
In the bottom panel the same velocity vectors as in the top panel are shown but now corrected for the LOS velocity of M31, i.e., in the rest-frame of M31 assuming no considerable perpendicular velocity (proper motion) for Andromeda \citep{kahn59,einas82}, although there are arguments which allow for a significant proper motion component of M31 \citep{loeb05}. The velocity component $\vect{v}_{\rm T,LOS}$ perpendicular to the line-of-sight of a satellite galaxy is unknown. Therefore a whole family of velocity vectors is possible for the satellite, illustrated by dotted vectors. 

The transformation of velocity vectors into the Andromeda-centric rest-frame is calculated by taking the time derivative of Eq.~(\ref{eqn_generaltransformation}):
\begin{equation}
\vect{v}_{\rm M31} = \matr{R}_{\rm M31} \left( \vect{v}-\vect{v}(\mbox{\small M31}) \right)\;.
\end{equation}
Upon setting $\vect{v}(\mbox{\small M31}) \equiv \vect{v}_{\rm LOS}(\mbox{\small M31})$ the line-of-sight velocity vectors of the satellites are calculated in the Andromeda-centric coordinate system. The LOS velocity vector of the satellite can now be split into a radial and a perpendicular component, $\vect{v}_r$ and $\vect{v}_t$, respectively. The values of the radial and perpendicular velocity components relative to Andromeda are given in Table~\ref{tab_M31data} for the M31 satellites, negative radial velocity meaning that the component is pointing towards M31, positive that it is pointing away.

This information is used next to set some limits on the possible poles of the orbits of the M31 satellites. In the example shown in Fig.~\ref{fig_m31vel} without an additional velocity component perpendicular to the line-of-sight, the sense of rotation about M31 would be counterclockwise (within the MW--M31--satellite plane). Only a large enough velocity component $v_1$ perpendicular to the LOS and within the MW--M31--satellite plane can reverse the sense of rotation. Any component $v_{\rm T,LOS}$ perpendicular to the MW--M31--satellite plane cannot reverse the sense of rotation, but can only displace the direction of the angular momentum vector along the polar path by a maximum $\pm 90\degr$ from the direction derived assuming no perpendicular velocity.

We have to make some additional assumptions about the maximum allowed velocities of the satellites. We assume the satellites are bound to an isothermal dark-matter halo of M31 with circular velocity $v_c \approx 250\unit{km\;s^{-1}}$, truncated at $r_c =250\unit{kpc}$. The escape velocity is given by

\begin{equation}
v_e(r) = \left\{
  \begin{array}{l@{\quad:\quad}l}
    \sqrt{2} \, v_c \sqrt{1+\ln r_c - \ln r} & r<r_c \\
    \sqrt{2} \, v_c \sqrt{r_c/r}     & r \ge r_c
  \end{array} \right. . \label{eqn_ve}
\end{equation}

A velocity component perpendicular to the LOS in the plane of the MW, M31, and the satellite can either contribute to the perpendicular velocity (giving a total velocity vector $\vect{v}_2$) or counteract ($\vect{v}_1$). We calculate the maximum possible velocity vectors $\vect{v}_1$ and $\vect{v}_2$ in the M31 rest-frame with absolute values of the velocities $|\vect{v}_1| = |\vect{v}_2| = v_e(r_{\rm s})$ (Eqn.~\ref{eqn_ve}), where $r_{\rm s}$ is the distance of a satellite from the centre of M31. The direction of the orbit of a satellite can now be restricted if the perpendicular components of $\vect{v}_1$ and $\vect{v}_2$ \emph{both} point in the same direction, i.e., if
\begin{equation}
s_v = \frac{ \boldsymbol{v}_1 \cdot \boldsymbol{v}_t }{ |\boldsymbol{v}_1||\boldsymbol{v}_t| }\;
\frac{ \boldsymbol{v}_2 \cdot \boldsymbol{v}_t }{ |\boldsymbol{v}_2||\boldsymbol{v}_t|}
  \equiv +1
\end{equation}
where `$\cdot$' denotes the scalar product of vectors.
This allows restriction of the polar paths to at least arcs of $180\degr$ which we call ``restricted polar paths'' (RPPs), i.e., the orientation of the orbit about M31 can be inferred.

\subsection{Application of restricted polar paths}\label{sec_rppapp}
\begin{figure*}
  \resizebox{16cm}{!}{
    \includegraphics{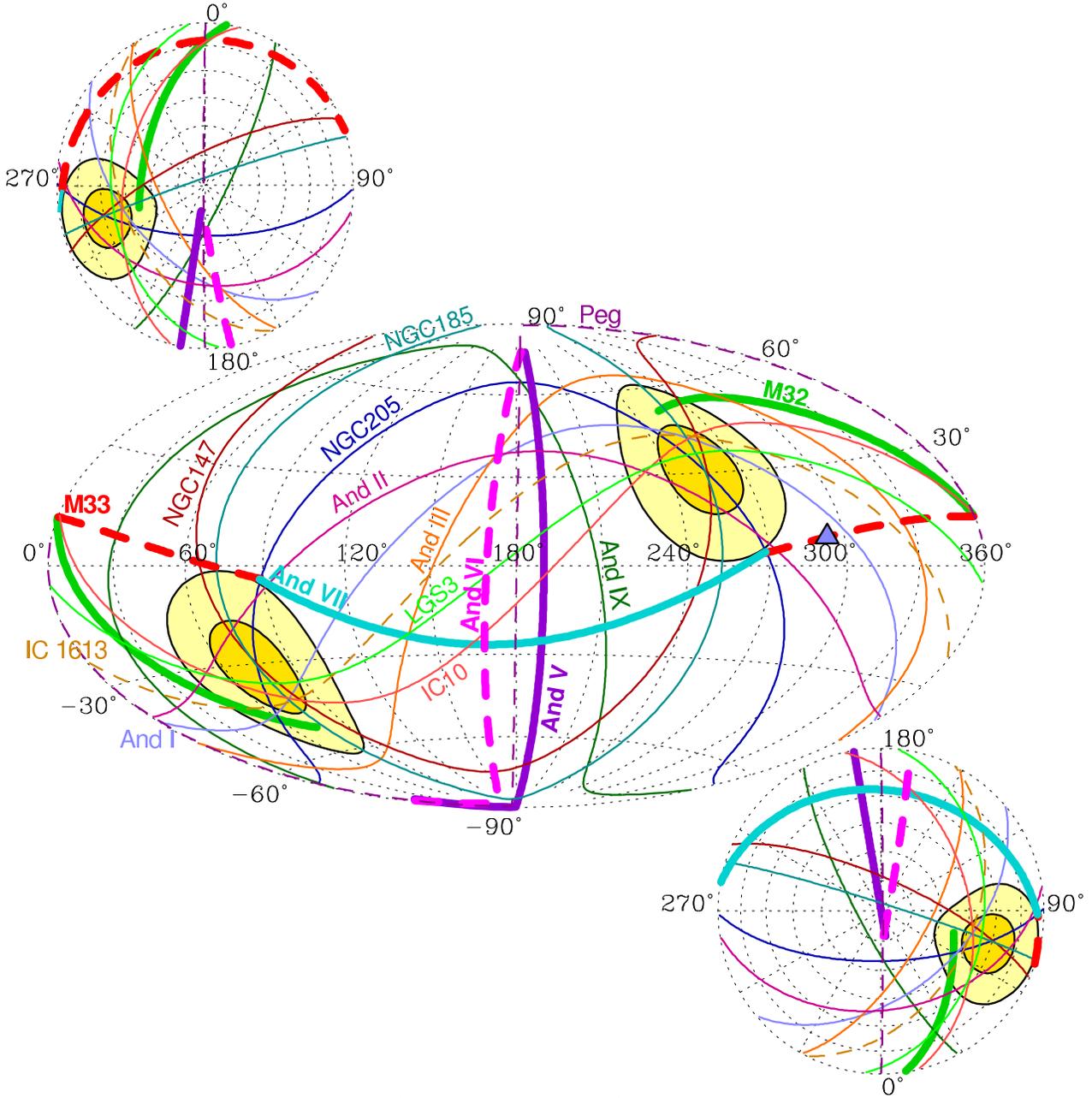}
  }
  \caption{Polar paths and restricted polar paths (marked with thicker lines) of the Andromeda satellite galaxies for the MI data-set. An overall view is shown using an Aitoff-projection. The top-left and bottom-right plots show a Lambert-projection of the northern and southern hemisphere, respectively. Solid paths are for the innermost dozen satellites entering the plane-fitting (\S~\ref{sec_analysisM31}) while dashed paths are the polar paths for those satellites not used. The direction of the angular momentum vector of M33, as derived from measured radial velocities and proper motion, is marked by the triangle. Regions of $15\degr$ and $30\degr$ distance from the fitted pole (Table~\ref{tab_fit}, entry `ALS') are indicated by the closed, thick solid loops.}
  \label{fig_m31gcpfMI}
\end{figure*}
\begin{figure*}
  \resizebox{16cm}{!}{
    \includegraphics{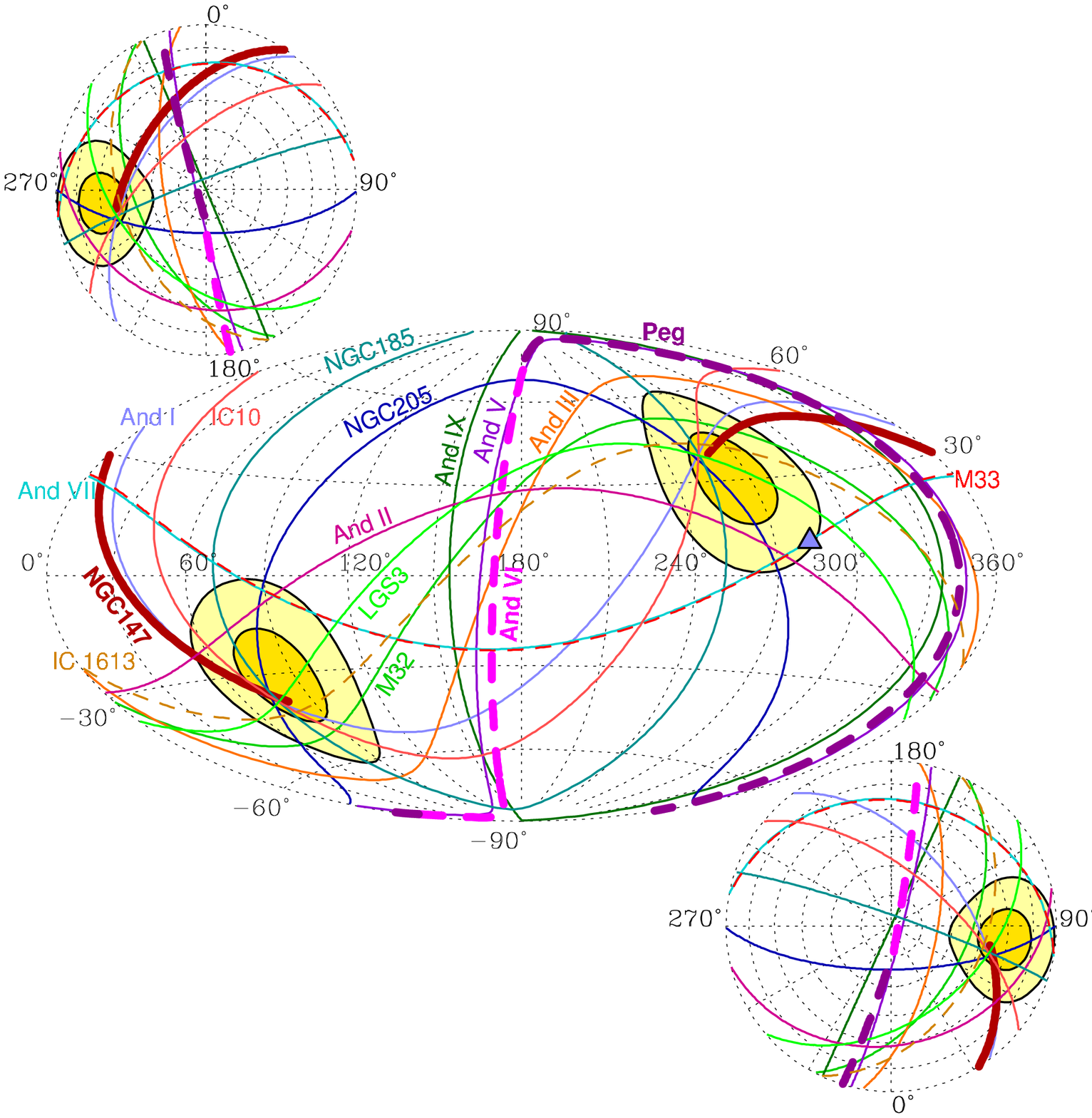}
  }
  \caption{As Fig.~\ref{fig_m31gcpfMI} for the KG data-set.}
  \label{fig_m31gcpfKG}
\end{figure*}
RPPs are calculated using heliocentric radial velocities as given in \citet[their table~1]{mccon06} for both data-sets. In Table~\ref{tab_M31data} we mark the data for those satellites with an asterisk for which we can restrict the poles by the RPP method described above. Only for And~VI the poles can be restricted for both data-sets. The polar paths and the resulting RPPs are plotted in Fig.~\ref{fig_m31gcpfMI} for the MI data and in Fig. \ref{fig_m31gcpfKG} for the KG data. Three projections for each data-set are displayed: an Aitoff-projection shows the overall global distribution of the polar paths, while it is difficult to distinguish the different tracks near the poles. For this reason two Lambert-projections are plotted in addition, in the top-left for the northern hemisphere, and in the bottom-right for the southern hemisphere. Data for the different satellites are plotted in different colours (see online material for coloured version). The polar paths for those satellites which were used for fitting the plane in Sect.~\ref{sec_analysisM31} are plotted with solid lines (satellites 1 -- 12 in Table~\ref{tab_M31data}), for satellites not used in the fitting with dashed lines. The position of the angular momentum vector of M33 as derived from the measured radial velocity and proper motion \citep{brunt05} is marked by a triangle assuming no proper motion for Andromeda. A considerable proper motion component of M31 would shift the direction of the angular momentum vector of M33 along its polar path in Figs.~\ref{fig_m31gcpfMI} and \ref{fig_m31gcpfKG} since the measured velocity of M33 is transformed to the rest-frame of M31. In addition we plot loops with distances of $15\degr$ and $30\degr$ from the pole of the fitted plane using the ALS method for all satellites within the approximate virial radius of M31 (\S\ref{sec_analysisM31}). Since the direction of the normal is arbitrary for the fitted plane, two regions appear on opposite sides of the sphere.

There is a strong clustering of intersection points of polar paths near the pole of the fitted plane, both in the northern and southern hemisphere. Those satellites whose polar paths intersect there are candidates for kinematic streams. In particular these are: NGC~205, And~I, NGC~147, NGC~185, IC~10, LGS~3, and IC~1613 (though not included in fitting the plane) within $15\degr$ from the pole for both data-sets. Within the $30\degr$ region, we additionally identify M32 and And~II. For the MI data, And~II is also found within the $15\degr$ region. \emph{Both data-sets, MI and KG, thus suggest the same satellites to be members of a possible stream}, based on the intersection of their possible kinematical poles coinciding with the direction of the normal to the fitted plane. If this clustering is true the satellite disc would be rotationally supported.
The nine satellites are exactly those which \citet{mccon06} identified as members of the possible candidate streams, namely (iii), (iv), and (v) in their paper. In total they identified five further possible streams. However, the distance uncertainties for Andromeda and its satellite galaxies are so large that an identification of possible streams based on the intersection-points alone may not be sufficient. With the additional argument of a spatially plane-like distributed satellite sample a stronger hint for possible streams emerges.

And~VI, which was excluded from the plane fitting as an outlier, is the only satellite for which the polar paths can be restricted in both data-sets. The resulting RPP is nearly perfectly a meridian at $l_{\rm M31} \approx 170\degr$. Thus, for And~VI the pole of the angular momentum vector is far off the pole of the fitted plane, and the orbit of the satellite can not be located within the plane. For the MI data-set the orbit of And~V can also be restricted and the resulting RPP is very close to that of And~VI. This may be an indication for a common direction of the angular momentum vectors of And~V and VI.

M33 is the only Andromeda companion for which a measured proper motion is available \citep{brunt05}. While M33 is located spatially close to the disc of the morphological subsample ({\sf mss8}, \S\ref{sec_mss} \& Fig.~\ref{fig_kgsss}), it cannot orbit within this disc: if this disc is rotationally supported, then the direction of all orbital angular momentum vectors must be close to $l_{\rm M31}=168\degr$, $b_{\rm M31}=-27\degr$: we restricted the polar paths of And~VI (and also And~V in the MI data), members of the {\sf mss8} subsample, to this area on the Andromeda sky. The direction of the M33 pole is far off this position ($>120\degr$), leading to the conclusion that M33 and the {\sf mss8} subsample of satellites cannot have a common orbital plane.\\

As a caveat we note that the RPP criterion applied above to the polar paths depends on two uncertain matters: the relative distance uncertainties and the true proper motion of Andromeda. Thus we can only treat the restriction criterion as a hint for more plausible regions of polar paths. Also note that a plane fitting algorithm and the appearance of a clustering of intersection points are not independent. The intersection point of the polar paths of two satellites is the direction of the normal of the plane containing these two satellites and the coordinate origin.

\subsection{Kinematic subsample of Andromeda satellites}\label{sec_kss}
Applying ALS to fit a plane for the kinematically motivated subsample of nine satellites given above (M32, NGC~205, And~I, NGC~147, And~II, NGC~185, IC~10, LGS~3, and IC~1613), but for the time being excluding IC~1613, the pole comes out to be located at $(l_{\rm M31}=69.9\degr ,\, b_{\rm M31}=-35.2\degr)$ [$(l_{\rm M31}=73.6\degr ,\, b_{\rm M31}=-35.0\degr)$] (MI[KG]-data-set), with axis ratios $c/a=0.12$ and $b/a=0.50$ [$c/a=0.15$ and $b/a=0.71$], i.e., \emph{a highly oblate (disc-like) configuration in both data-sets} (Table~\ref{tab_fit}). The pole is very close to the pole found for the full sample of twelve satellites within the approximate virial radius. In Fig.~\ref{fig_m31fit12} it is clearly visible that also some satellites are excluded here (encircled by light gray circles) which lie spatially close to the initially fitted plane.
Including the very distant (and possibly bound) satellite IC~1613 we find the pole of the fitted normal to be located at $(l_{\rm M31}=74.4\degr ,\, b_{\rm M31}=-40.5\degr)$ [$(l_{\rm M31}=74.7\degr ,\, b_{\rm M31}=-40.6\degr)$], with axis ratios $c/a=0.10$ and $b/a=0.35$ [$c/a=0.11$ and $b/a=0.43$]. The latter axis ratio, leading to a more prolate configuration, is totally dominated by this one very distant satellite (IC~1613). 
We concentrate our further analysis on the sample of eight satellites without IC~1613, because it is outside the approximate virial radius of M31. We refer to this subsample in Tables \ref{tab_fit} -- \ref{tab_resbs} as {\sf kss8} (kinematic subsample of eight satellites).

Applying the AE test to the {\sf kss8} subsample the direction of the principal axis is $(l_{\rm M31}=70.5\degr ,\, b_{\rm M31}=-35.2\degr)$ [$(l_{\rm M31}=74.2\degr ,\, b_{\rm M31}=-35.2\degr)$] with a spherical standard distance $\Delta_{\rm sph}=2.4\degr$ [$\Delta_{\rm sph}=1.6\degr$] (KG [MI] data). The location of the principal axis is in good agreement with the single fits above, and $\Delta_{\rm sph}$ is for both data-sets significantly smaller than for the full data-sets, and also an order of magnitude smaller than for the {\sf mss8} subsample (\S~\ref{sec_mss}).

Performing the bootstrap analysis with 5\,000 re-samplings for the kinematically motivated subsample yields shape parameter $\gamma=5.9$ and strength parameter $\zeta=4.7$ [$\gamma=14.9$, $\zeta=4.2$]. For both data-sets the distribution of the directions of the fitted planes of the bootstrapped data are strongly concentrated and clustered. The derived spherical standard distance is $\Delta_{\rm sph}=9.8$ [$\Delta_{\rm sph}=11.5$], indicating that the systematic effects caused by the distance uncertainties are smaller than the internal scatter as derived by the bootstrapping.

Even though the morphologically motivated subsample (\S~\ref{sec_mss}) has a smaller rms-height $\Delta$ than the kinematically motivated subsample for the KG-data and thus appearing as `thinner' disc, the bootstrap analysis shows that the latter one has a more pronounced planar-like feature. The strong clustering is found in both data-sets. \emph{We have therefore uncovered a sample of eight M31 satellites (nine if IC~1613 were included) which span a very pronounced disc-of-satellites that is probably rotationally supported.}

\section{Statistical significance of disc-like distributions}\label{sec_bootstrap}
In order to study the possible physical nature of the MW satellite system the statistical significance of the observed anisotropy given a parent distribution needs to be quantified. According to the null-hypothesis, the parent distribution ought to be a dark-matter sub-halo distribution which may be spherical, oblate, or prolate.  To evaluate the significance of planar distributed satellite systems we compare the bootstrapped samples of the observed distribution with bootstrapped data of random samples from the parent distribution. For this we first create spherically isotropic distributions, where the radial linear probability density is proportional to $\rho(r) \propto r^{-p}$, $p=2$ ($\Rightarrow \rho_{\rm sph}(r,\vartheta,\phi) \propto r^{-q}$, $q=4$) consistent with the radial distribution found for the Milky Way \citep{kroup05} and Andromeda \citep{koch06}. Random oblate, prolate, or triaxial ellipsoidal distributions with axis ratios $c/a$ and $b/a$ are then constructed by scaling the components of the random spatial position vectors while keeping the volume of the ellipsoid invariant. As shown in Eq.~(\ref{eqn_rmsD}) the formally expected relative height of a spherical distribution is dependent on the minimum and maximum radii. Therefore the random samples are set-up with the minimum and maximum radii as found for the Milky Way (see Table~\ref{tab_MWdata}). For ellipsoidally distributed random samples the initial values are scaled such that the final distribution has the expected minimum and maximum radii.

\begin{figure}
  \resizebox{\hsize}{!}{
    \includegraphics{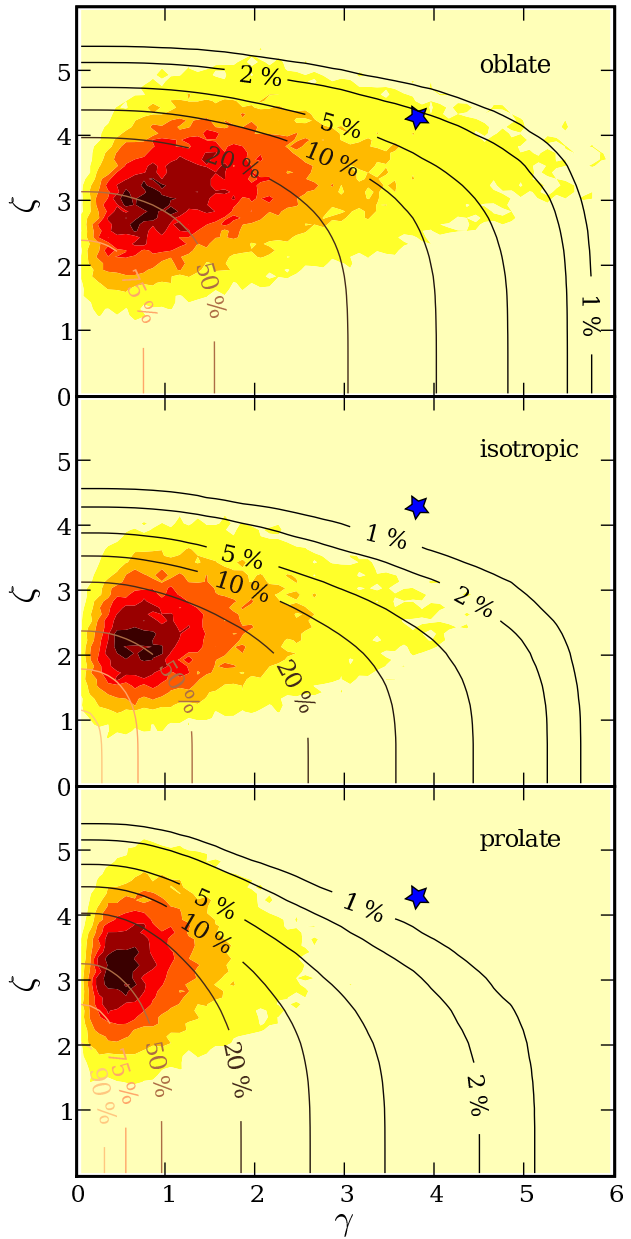}
  }
  \caption{A contour-plot of strength parameter $\zeta$ versus shape parameter $\gamma$ for 100\,000 random samples each consisting of eleven model satellites with an isotropic (central panel), oblate ($c/a=0.5$, $b/a=1.0$, top panel), and prolate ($c/a=b/a=0.5$, bottom panel) parent distribution, each individually bootstrapped 10\,000 times. The shaded contours show the density distribution of the derived parameters, dark being high density. The contourlines show the enclosed values with significance levels as labelled. The star marks the parameters derived for the MW.} \label{fig_gammazeta}
\end{figure}

As for the observed data, each random sample is bootstrapped 10\,000 times and we calculate the shape parameter $\gamma$ and the strength parameter $\zeta$ of the resulting distribution of fitted normal vectors. Fig.~\ref{fig_gammazeta} (central panel) shows a contour-plot $\zeta$ vs. $\gamma$ derived for 100\,000 random samples from an isotropic distribution ($a=b=c$) each consisting of eleven model satellites, bootstrapped $10^4$ times. As can be seen from the shaded regions, which show the density distribution, the distribution of normal vectors of bootstrapped random samples is not expected to be randomly distributed in $\gamma$, $\zeta$ space. They are typically found to be girdled or transitional ($\gamma \la 1$) and marginally concentrated ($\zeta<3$), while there is also some fraction of clustered ($\gamma > 1$) distributions.
For oblate parent configurations (Fig.~\ref{fig_gammazeta}, upper panel) much more clustered distributions ($\gamma>1$) result which are typically also more concentrated ($\zeta \ga 3$). On the other hand, for prolate parent configurations (Fig.~\ref{fig_gammazeta}, bottom panel) typically a much higher fraction of girdled distributions ($\gamma<1$) results, but also with higher concentration parameter ($\zeta \ga 3$). Small concentration parameters are mostly found for an isotropic distribution.

To calculate the significance of an observed distribution the joint distribution function $D(\gamma,\zeta)$ is computed and the percentile of bootstrapped random samples is derived for which both, the shape parameter and the strength parameter, are larger than found for an observed satellite distribution, e.g. of the Milky Way (Fig.\ \ref{fig_gammazeta}, contour lines). For each parameter pair of initial values $c/a$ and $b/a$ we create 100\,000 random satellite samples each consisting of eleven satellites. Each of these samples is individually analysed using the bootstrap method with 10\,000 re-samplings. This required a large amount of CPU power and we ran the simulation on the computer system of the Argelander-Institute and the CIP-pool\footnote{\url{http://cip.physik.uni-bonn.de}} of the physics department. The full run took about 7\,500 CPU-hours running simultaneously on up to 30 PCs from 500~MHz class to 3~GHz class CPUs using a distributed computing technique.

\subsection{The Milky Way}\label{sec_statsigMW}
\begin{figure}
  \resizebox{\hsize}{!}{
    \includegraphics{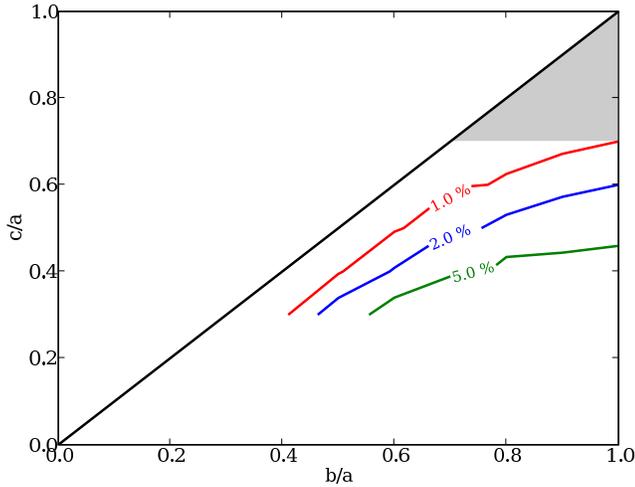}
  }
  \caption{Approximate contour lines for 1, 2, and 5~per cent probability that the innermost eleven Milky Way satellites are drawn randomly from a parent distribution with initial axis-ratios $c/a$ and $b/a$. The grey shaded region shows the range of axis ratios typically found for dark-matter haloes in numerical simulations.}
  \label{fig_statsig}
\end{figure}

The percentile of models found to have bootstrapped distributions more concentrated than for the Milky Way are listed in Table~\ref{tab_rndbootmw}. Fig.~\ref{fig_statsig} shows approximate contour lines for 1, 2, and 5~per cent probability that the MW satellites (KTB data-set) are drawn randomly from a parent distribution with initial axis-ratios $c/a$ and $b/a$. The values typically obtained for Milky Way sized DM haloes \citep[see e.g.,][]{libes05} are shown by the grey shaded region.

\begin{table}
\caption{Percentile of bootstrapped random samples for which the shape parameter and the strength parameter indicate a more concentrated distribution of the normals of bootstrapped satellites than found for the Milky Way satellite system. Different setup distributions are used with axis ratios $c/a$ (along rows) and $b/a$ (along columns). The top table gives the percentile for the innermost eleven satellites, the bottom table for the innermost twelve satellites including the UMa dwarf satellite candidate. Parameter combinations for which $c/b<b/a$, i.e. which are triaxial and more oblate, are highlighted by a light grey background colour.}\label{tab_rndbootmw}
\centering
\begin{tabular}{ccccccccc}
\hline
 & \multicolumn{6}{c}{$b/a$} \\
    $c/a$ & {\it 0.3} & {\it 0.4} & {\it 0.5}  & {\it 0.6}  & {\it 0.7}  & {\it 0.8}  & {\it 0.9}  & {\it 1.0} \\
{\it 1.0} &     &     &     &           &            &           &           &      0.5  \\
{\it 0.9} &     &     &     &           &            &           &      0.5  & \sgcc{0.5} \\
{\it 0.8} &     &     &     &           &            &      0.6  & \sgcc{0.6} & \sgcc{0.6} \\ 
{\it 0.7} &     &     &     &           &       0.6  &      0.7  & \sgcc{0.8} & \sgcc{1.0} \\
{\it 0.6} &     &     &     &      0.7  &       0.8  & \sgcc{1.1} & \sgcc{1.5} & \sgcc{2.0} \\
{\it 0.5} &     &     & 0.7 &      0.9  &       1.5  & \sgcc{2.4} & \sgcc{3.3} & \sgcc{4.1} \\
{\it 0.4} &     & 0.5 & 0.9 &      2.1  & \sgcc{4.1} & \sgcc{6.3} &          &           \\
{\it 0.3} & 0.2 & 0.8 & 2.7 & \sgcc{6.9} & \sgcc{12.2} &           &           &           \\
\hline
{\it 1.0} &     &     &     &           &            &           &           &      0.3  \\
{\it 0.9} &     &     &     &           &            &           &      0.3  & \sgcc{0.4} \\
{\it 0.8} &     &     &     &           &            &      0.4  & \sgcc{0.3} & \sgcc{0.5} \\
{\it 0.7} &     &     &     &           &       0.5  &      0.4  & \sgcc{0.6} & \sgcc{0.6} \\
{\it 0.6} &     &     &     &      0.5  &       0.5  & \sgcc{0.8} & \sgcc{1.1} & \sgcc{1.5} \\
{\it 0.5} &     &     & 0.4 &      0.7  &       1.1  & \sgcc{2.0} & \sgcc{2.7} & \sgcc{3.6} \\
{\it 0.4} &     & 0.3 & 0.7 &      1.6  & \sgcc{ 5.4} & \sgcc{3.2} &           &           \\
{\it 0.3} & 0.1 & 0.6 & 2.2 & \sgcc{6.2} & \sgcc{11.6} &           &           &           \\
\hline
\end{tabular}
\end{table}

The same analysis for a sample of twelve satellites is repeated with 20\,000 random samples and the derived shape parameters are compared with those found for the Milky Way satellites including the UMa dwarf galaxy. The resulting fractions are listed in the bottom part of Table~\ref{tab_rndbootmw}.
Including the most recently discovered dSph in Canes Venatici we ran only one test with 13 satellites and 20\,000 random samples for a spherical setup. For this run we find 0.3~per cent of the random samples more concentrated than for the Milky Way sample.

The null-hypothesis that the Milky Way satellites are drawn randomly from a spherical or mildly triaxial parent population can be excluded at very high statistical significance ($ \ge 99.5$~per cent, confirming the results of \citealp{kroup05}). 
With increasing triaxiality, the probability increases for oblate configurations (which are marked with a grey background in Table~\ref{tab_rndbootmw}). Prolate configurations are basically excluded, except for configurations nearly perfectly triaxial, e.g. for $c/a=0.5$ and $b/a=0.7$, where the probability may be of order 1~per cent. Including the UMa dwarf galaxy increases the significance of this result (reduces the propability).

\subsection{Andromeda}\label{sec_statsigM31}
For Andromeda, using an appropriate setup, the probability that the satellite distribution is drawn randomly from a spherically isotropic parent distribution is already 12~per cent for the KG-data. This reflects the fact that we find the bootstrapped normals of the M31 satellite system to be much less clustered than for the Milky Way. The distribution of bootstrapped normals for the MI data-set are even less clustered and the probability is thus larger. So the hypothesis that the M31 satellite system is drawn randomly from a spherically isotropic parent distribution can not be rejected at present, using the available data.

\subsection{Statistical significance of M31 subsamples}\label{sec_statsigM31ss}
Performing the analysis as described above for the morphologically (\S\ref{sec_mss}) and kinematically (\S\ref{sec_kss}) motivated subsamples would \emph{not} yield the correct significance: the null-hypothesis for deriving the statistical significance is that all satellite galaxies within a certain radius trace a parent distribution, i.e. one assumes all satellite galaxies to be of the same origin, namely luminous DM sub-structures. But performing the analysis as above would imply that the eight satellites selected by the morphological or kinematical criterion are luminous DM sub-structures and the excluded ones are of another origin. Even in the case of the morphologically motivated subsample, where one may argue that all dSphs/dEs have been build due to the same mechanism, some were excluded because of their large distance from the fitted plane. But they should have been included to derive the significance, because otherwise one implies a different origin of the excluded satellites. \citet{koch06} did not take this into account.

To derive the significance correctly, the procedure is to set up twelve model satellites within the appropriate distance range from M31 and then select all combinations of eight model galaxies out of the full sample. For each of these subsamples a full bootstrap analysis has to be performed. Then the fraction of random samples that have subsamples of eight model satellites with a more pronounced disc-like distribution than the observed sample needs to be calculated.

Practically, this analysis can only be performed for a few specific set-ups because the runs take a large amount of CPU time. We performed one run for each of the data-sets, MI and KG, respectively, using an isotropic parent distribution. The differences between both random runs are the minimum and maximum radii, chosen to match the values of the corresponding observed data-sets. We set-up 10\,000 random samples. The full run took about 20\,000 CPU hours, running simultaneously on more than 30 standard PCs.

The probability that the \emph{morphologically} motivated subsample is picked from samples that are randomly drawn from an isotropic distribution is 100~per cent for the MI data-set and 95~per cent for the KG data-set. The probabilities for the \emph{kinematically} motivated subsample are 17~per cent for the MI data-set and 10~per cent for the KG data-set. Even though appearing as a `thin' disc of satellites, we conclude that the morphologically motivated subsample is just a chance alignment of galaxies. For the {\sf kss8} subsample we can also not reject the hypothesis that the satellites are picked from a random sample, but the probability is much lower ($<17\%$) for both data-sets.

\section{Concluding remarks}\label{sec_discussion}
We introduce a framework of mathematical methods to analyse the spatial distribution of the satellite galaxies of the Milky Way and Andromeda. A statistical analysis based on the `thickness' of a planar-like distribution alone is not sufficient to characterise the distribution \citep{kang05,zentn05}. The bootstrap method is used to derive the distribution of poles of fitted planes which are analysed using methods based on the statistical analysis of spherical data. These methods quantify the robustness of a planar-like distribution. Thus, a population of satellites that is not planar-like will in our analysis be robustly  identified as an unstable distribution of poles. We make our implementations of the algorithms available under an open source licence at \url{http://www.astro.uni-bonn.de/downloads}. Note that the analysis presented here is based on more than three CPU years on state-of-the-art PCs using a distributed computing technique.

\subsection{The Milky Way system}
Applying two methods (ALS and ODR) to fit planes, the Milky Way satellite system within $254\unit{kpc}$ is found to be highly anisotropical. All companion galaxies are aligned in a disc-like structure with a rms-height of only $\Delta=18.5\unit{kpc}$ -- $22.8\unit{kpc}$ (Table~\ref{tab_fit}). This disc-of-satellites is highly inclined with respect to the MW disc, $|b_{\rm MW}| \approx 12\degr$, passing the Galactic plane close to the Galactic Centre, $D_{\rm P} \approx 8\unit{kpc}<\Delta$.

Satellite galaxies are believed to be luminous dark matter subhaloes. In recent cosmological simulations MW sized DM haloes are found to be triaxial, $c/a=0.6$--$0.8$, typically more prolate than oblate \citep[e.g.][]{jing02,bullo02,libes05,zentn05}. \citet{libes05} showed that the distribution of a large number of dark-matter subhaloes within these host haloes has a similar shape, i.e. the distribution of dark-matter subhaloes is a fair tracer of their host halo. This distribution, the large sample of subhaloes within the host halo, is the parent distribution of the Milky Way satellites \emph{if} the satellite galaxies are luminous, dark matter dominated subhaloes.

Applying our new statistical framework to the satellite system of the Milky Way we show that the hypothesis that the MW satellites are drawn randomly from an isotropic or mildly triaxial distribution can be excluded at a high statistical significance level ($\ge 99.5$~per cent). It can also be excluded that the parent distribution has a prolate shape as derived from CDM theory. The null-hypothesis that the satellite system is drawn randomly from a dark-matter parent distribution cannot be rejected \emph{only} if the parent distribution is highly triaxial and oblate, i.e. if the parent distribution is already disc-like. In this case, and as long as no host-galaxy formation is included in large scale CDM simulations, the disc of the Milky Way has to be \emph{postulated} to be nearly perpendicularly oriented to the highly oblate host halo, because we find the disc-like structure of satellite galaxies to have a polar alignment. Even so, the required highly triaxial oblate DM-host shape is not consistent with the results of modern CDM structure formation simulations. Furthermore, recent measurements of the shape of the MW potential using the Sagittarius stream show it to be spherical within about $60\unit{kpc}$ \citep{fellh06}.

The recent discoveries of two additional faint MW companions increases the confidence of the above statements. It therefore follows that any $\Lambda$CDM sub-structure distribution is inconsistent with the observed morphology of the MW satellite population (Fig.~\ref{fig_statsig}). However, since the SDSS \citep{york00} mostly covers the north pole region of the Galactic sky (Fig.~\ref{fig_satlbmw}), newly detected dwarf galaxies are very likely close to the fitted disc. To get an answer beyond this possible bias it will be crucial to extend the search for MW companions over a larger area of the sky and particularly at lower galactic latitudes as is planned by the Stromlo Missing Satellite Survey (Jerjen et al. 2006, in preparation) using the new ANU SkyMapper telescope \citep{schmi05}. We note though that if additional very faint dwarf galaxies are discovered to not lie within the great disc of satellite galaxies as quantified here, we are nevertheless left with the fact that the eleven most luminous MW satellite galaxies are aligned in the disc.

For the Milky Way a significant fraction of dwarf galaxies may be invisible in the optical due to obscuration by the Galactic disc at low galactic latitudes \citep[e.g.,][]{mateo98}. Within the virial radius of the Milky Way about half of the total volume has latitude $b \le 30\degr$. Andromeda is in that sense a better probe since it's halo is not that much affected by obscuration \citep{mccon06}. A simple estimate for the MW, assuming that all undetected satellites with $b \le 15\degr$ are obscured, 50~per cent of all undetected satellites with $15\degr < b \le 30\degr$ are obscured, and assuming that the undetected satellites are homogeneously distributed over the whole sky, we find that about 35~per cent of all satellites with $b \le 30\degr$ may be found more than $1\sigma=\Delta$ and about 30~per cent more than $3\sigma$ off the fitted disc-of-satellites.

\subsection{The Andromeda system}
For the Andromeda satellite system it cannot be excluded that it has been drawn randomly from a spherical isotropic parent distribution. However, we do find the M31 satellite system to be anisotropic, but the details depend on the data-set used. The fitted disc-like structure for all satellites within the approximate virial radius is not as polar-aligned as for the MW ($|b_{\rm M31}|\approx30\degr$) and it is approximately twice as `thick' as found for the MW (Table~\ref{tab_fit}).

Two incompatible subsamples of satellite galaxies which have a disc-like distribution can be identified: one morphologically motivated as proposed by \citet{koch06} and one kinematically motivated (\S\ref{sec_kss}). Since the disc-like satellite system of the MW is dominated by dSph galaxies, one can speculate about a common building mechanism for all the dSphs in the Local Group. If this mechanism was the break-up of a large, gas-rich galaxy or the formation of tidal-dwarf galaxies (TDGs) in an early major-merger event, one expects the dSphs to have initially correlated directions of their angular momentum vectors, supporting a disc-like structure. This would favour the morphologically motivated subsample since it is initially build up of dSphs only. However, at least two of the Andromeda dwarf spheroidals within the virial radius do not fit into this picture and have to be excluded because of their apparently large distance from the plane. Also only one out of three morphologically similar dEs is found close to the disc. But we cannot exclude the possibility that some massive TDGs may retain their interstellar media to appear today as dIrr galaxies \citep{hunte00,recch06a}.

The disc-like distribution of the dSph/dE galaxies around Andromeda is present for the KG data-set only, but can not be identified for the MI data-set. Comparing the results of the AE and the bootstrap test it follows that even in the KG data-set the systematic errors caused by the distance uncertainties are larger than the intrinsic scatter as quantified by the bootstrapping. The high significance of the morphologically motivated disc as derived by \citet{koch06} is incorrect because of three reasons: firstly they derived their significance based on the thickness of the disc alone, secondly they used the ODR fitting method which is affected by the systematic alignments of the distance uncertainties of the M31 satellites, and thirdly they did not use the correct null-hypothesis to derive the significance. We find the distribution of the morphologically motivated subsample of M31 satellite galaxies to be fully consistent with being picked from a random distribution.

Combining the polar-paths of the Andromeda satellites with the fitted plane it follows that there may be a kinematic association of some of the M31 satellites: M32, NGC~205, And~I, NGC~147, And~II, NGC~185, IC~10, LGS~3, and probably, though due to its large distance unlikely, IC~1613. These satellite galaxies were also identified as possible stream members by \citet{mccon06}, while there are also other possible streams which now seem to be less likely. This sample of eight Andromeda satellites forms a very pronounced thin disc with inclination of $\approx 31\degr$ away from a polar alignment, and that holds true for both data-sets (MI and KG). The kinematically motivated subsample is found to have a much more pronounced disc-like distribution than the morphologically motivated one, although its thickness is larger. Even though the kinematically motivated subsample has a much higher statistical significance than the morphologically motivated one, we can also not exclude the possibility that this sample is picked from a random distribution.

We argue that the close proximity of M33 to the disc of the morphologically motivated subsample is a pure chance alignment since the direction of its angular momentum vector can not be correlated with that of the M31 dSphs. Using the RPPs (\S\ref{sec_rpp}) we show that, if the dSphs have a common direction of their angular momentum vectors, it is far off ($\ga 120\degr$) the pole of the M33 orbit as derived from its measured proper motion. But M33's kinematical pole lies close to the pole of the whole distribution and to the pole of the kinematically motivated subsample (Figs.~\ref{fig_m31gcpfMI} and \ref{fig_m31gcpfKG}).

Using radial velocity measurements we created restricted polar paths, showing that at least the direction of the angular momentum vector of And~VI can be restricted to an arc segment of $180\degr$, i.e. the sense of rotation of the orbit around Andromeda can be restricted in both data-sets. For six of the M31 satellites the possible poles can only be restricted for one of the data-sets. Restricted polar paths hint to more likely regions of the locations of angular momentum vectors on the M31 sky.

The global influence of M32 and M33 on the satellite system of Andromeda remains to be studied. The Triangulum Galaxy M33 may be sufficiently massive \citep[$\ga 5 \times 10^{10} \unit{M_{\sun}}$,][]{herrm05} to mix up the satellite system of Andromeda or bring its own satellite galaxies into the M31 system. Similarly the cE M32 may have had a strong influence on the dynamical evolution of Andromeda's satellite galaxies if massive enough in the past \citep[e.g.,][]{bekki01}. 

\subsection{The Milky Way vs.\ Andromeda}
\begin{figure}
  \resizebox{\hsize}{!}{ \includegraphics{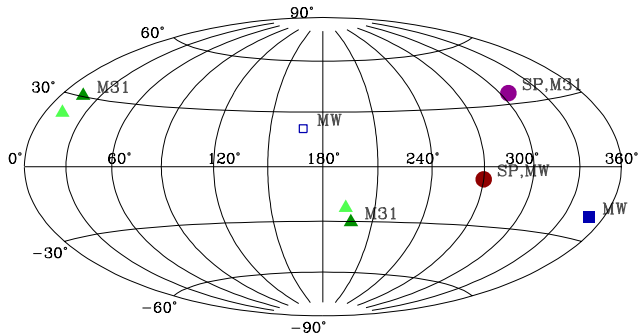} }
  \caption{An Aitoff projection in supergalactic coordinates of the directions of the spin poles of the MW and M31 discs marked by filled circles. The directions of the normals of the fitted plane of all satellites of the MW and M31 within their virial radii are marked by squares and triangles, respectively (MI data-set in dark grey, KG data-set in light grey).}
  \label{fig_supergalactic}
\end{figure}

There is no obvious evidence for a spatial association of the disc-like structure of the Milky Way satellites and that of the Andromeda satellites.
If M31 (located at $l_{\rm MW}=121.7\degr$, $b_{\rm MW}=-21.5\degr$) and its satellites were associated with the disc-of-satellites of the MW, M31 ought to be close to the fitted disc-like structure and the fitted planes ought to be aligned. Instead, M31 is $\approx 55\degr$ off the fitted plane of the MW satellites. Similarly, the angle between the normals of the fitted satellite planes of the MW and M31 is $\approx 50\degr$ -- $60\degr$, but would be $0\degr$ if they were perfectly aligned.

Fig.~\ref{fig_supergalactic} shows the directions of the spin-poles of the MW and M31, as well as the directions of the normals of the fitted disc-like structures of their satellite distributions in supergalactic coordinates \citep{devau91}. The directions of the spin-poles (SP) are marked by filled circles.
The directions of the normal of the fitted plane to the MW satellite system is marked by squares, the filled one pointing in the direction that is close to the kinematical poles as derived for some of the MW satellites (\citealt[and references therein]{palma02}; \citealt{kroup05}) indicating the sense of rotation if the disc-like structure is rotationally supported.
The direction of the fitted normal to the M31 satellite system marked by triangles is arbitrary (`up' or `down'). It follows that both spin-axes of the disc galaxies lie within about $30\degr$ of the supergalactic plane and that the disc-of-satellites of both hosts are likewise oriented such that their poles lie within $30\degr$ of the supergalactic plane. \emph{All} discs are thus highly inclined with respect to the supergalactic plane, but do not appear to be mutually aligned. This would appear to contradict the notion that the satellites accreted individually preferentially from the direction of the supergalactic plane, i.e. from the direction of the medium-scale ($\sim$ few Mpc) matter distribution.

The highly-inclined orientation of the stellar discs of the MW and of M31 relative to the supergalactic plane can be understood either as being a result of tidal torquing \citep{navar04} or resulting from the perpendicular collapse of matter onto the supergalactic plane (\citealp{doros73, doros78}; see e.g.\ also \citealp{hu98}). The disc-of-satellites of the MW and of M31 are both also highly inclined to the supergalactic plane \emph{and} the respective galactic discs of their hosts. Tidal torquing, if responsible for the orientation of the galactic discs, can therefore not be the origin of the disc-of-satellite orientations. Instead these two discs could also result from the collapse of matter onto the supergalactic plane. However, this would beg the question as to why they are so highly inclined to the host galactic discs. Alternatively, it would appear more natural or intuitive to understand the disc-of-satellites as being the result of stochastically occurring mergers which leave populations of related TDGs. Such populations would remain visible for highly inclined events relative to the host discs, because populations of TDGs in low-inclination orbits would precess apart and possibly end up merging altogether with their host discs \citep{penar02}.

\vspace{5mm}
\noindent{\bf Acknowledgements}\\
We thank Klaas S. de Boer, Michael Hilker, and Ole Marggraf for helpful comments, Jan Pflamm-Altenburg and Patrick Simon for useful discussions. We thank the anonymous referee for very helpful comments to improve this paper. This research has made use of the SIMBAD database, operated at CDS, Strasbourg, France.

\appendix
\section{Conditional bootstrapping}\label{sec_appbootstrap}
To uniquely perform a plane fitting, at least three different objects in a bootstrap sample are required. To correctly calculate the total number of possible distinct bootstrap samples, this can generally be formulated as:

\emph{What is the number of ways of picking $k$ unordered outcomes with replacement from $n$ possibilities under the condition that at least $q$ outcomes are different?}

Without the condition that at least $q$ outcomes are different, the total number is given by the number of unordered samples with replacement
\begin{equation}
N_{\rm tot} = {n+k-1 \choose k}\;.
\end{equation}

We start by calculating the number of ways of picking exactly $p=q$ different outcomes. This is given by the number of ways of picking $p$ unordered outcomes \emph{without} replacement from $n$ possibilities times the number of ways of picking $k-p$ unordered outcomes with replacement from $p$ possibilities,
\begin{equation}
N_{p=q} = {n \choose p} {p + (k-p) - 1 \choose k-p} = {n \choose p} {k - 1 \choose k-p} \; .
\label{eqn_condbootstrap}
\end{equation}
This can be understood as follows:
\begin{displaymath}
\overbrace{
  \underbrace{z_{j_1},z_{j_2}, \ldots ,z_{j_p}}_{p},\underbrace{z_{i_1},z_{i_2},\ldots,z_{i_{k-p}} }_{k-p}
}^{k}
\quad,\quad i_m \in \{j_1,j_2,\ldots,j_p\}.
\end{displaymath}
First choose exactly $p$ different outcomes from all $n$ possibilities $z_1,z_2, \ldots ,z_n$ without replacement (all $p$ must be different); the number of ways is given by ${n \choose p}$. Since exactly $p$ different outcomes were premised, it can now only be chosen from those $p$ outcomes that have been chosen in the first step (so now the number of possibilities is $p$), which are ${p + (k-p) - 1 \choose k-p}$ ways.

Next calculate this for $p=q+1$ and so on and finally add up all numbers of ways:
\begin{equation}
{^q}N_{\rm tot} = \sum_{p=q}^{k} {n \choose p} {k-1 \choose k-p}\;.
\end{equation}
Since the total number of possible distinct bootstrap samples is typically very large (e.g. ${^3}N_{\rm tot} = 77\,557\,275$ for $n=15$) a sufficiently large number of re-samplings is used, as is usually done in bootstrapping.

Note that bootstrapping is different to the sampling method \citet{koch06} employed, while misleadingly referring to it as bootstrap. These authors fitted a plane through every possible combination of $3, 4, \ldots, n$ objects, an approach that is related to the jackknife method. The different sampling method may have biased the results of \citet{koch06}.

\section{Extrasolar celestial coordinate systems}\label{sec_Eccs}
In order to study the three-dimensional distribution of a satellite system it is most convenient to transform their position vectors relative to the observer into an extrasolar coordinate system. We elaborate the transformation in much detail in the most general way, using vector and matrix operations, to allow for an as wide as possible range of applications. The algorithm described also allows for a free choice of the reference point of the $l=0\degr$ direction of the target system.
Similar transformations (in a less general way) were used in two recent relevant works on the spatial distribution of the Andromeda satellite galaxy system \citep{mccon06,koch06}, but the transformations differ significantly. As we will show the transformation as applied by \citet{koch06} is incorrect.

The basis system for the transformation may be any celestial coordinate system, e.g., equatorial coordinates ($\alpha$, $\delta$) or Galactic coordinates ($l$, $b$). The origin and the orientation of the target coordinate system must be given as (i) positional coordinates, (ii) inclination, and (iii) position angle in the basis system. We exemplify the transformation by transforming into an Andromeda-centric coordinate system.
In the following, bold symbols like $\vect{r}$ denote vectors, whereas $r$ is the absolute value of a vector $r=|\vect{r}|$. Unit vectors in a specific direction are denoted as $\vect{e}_r = \vect{r} / r$.

\subsection{An Andromeda-centric coordinate system}\label{sec_M31coords}
Let $(\vect{e}_{x}^\oAnd,\vect{e}_{y}^\oAnd,\vect{e}_{z}^\oAnd)$ be a Cartesian coordinate system centred on Andromeda similar to the galactocentric coordinate system $(\vect{e}_{x}^\oMW, \vect{e}_{y}^\oMW, \vect{e}_{z}^\oMW)$ of the Milky Way \citep[see also][]{mccon06}.
We define the coordinate system such that $\vect{e}_{z}^\oAnd$ is the direction of the normal to the stellar disc of Andromeda, $\vect{e}_{x}^\oAnd$ is the projected direction from the MW to M31 onto the disc of M31 and $\vect{e}_{y}^\oAnd=\vect{e}_{z}^\oAnd \times \vect{e}_{x}^\oAnd$.

Equatorial coordinates can be calculated as follows:
\begin{eqnarray}
l &=& \arctan \left( \frac{y}{x} \right) \label{eqn_lbr1}\\ 
b &=& \arcsin \left( \frac{z}{r} \right) \label{eqn_lbr2}\\
r &=& \sqrt{ x^2 + y^2 + z^2}            \label{eqn_lbr3} 
\end{eqnarray}
where $x,y,z$ are the components of a vector either given in the galactocentric or in the Andromeda-centric Cartesian coordinate system. Note that if we give galactocentric coordinates $(l_{\rm MW}, b_{\rm MW})$, the centre of the coordinate system is the Galactic Centre (GC), whereas the centre of the standard `Galactic coordinate system' $(l,b)$ is the Sun.

\begin{figure}
  \resizebox{\hsize}{!}{  \includegraphics{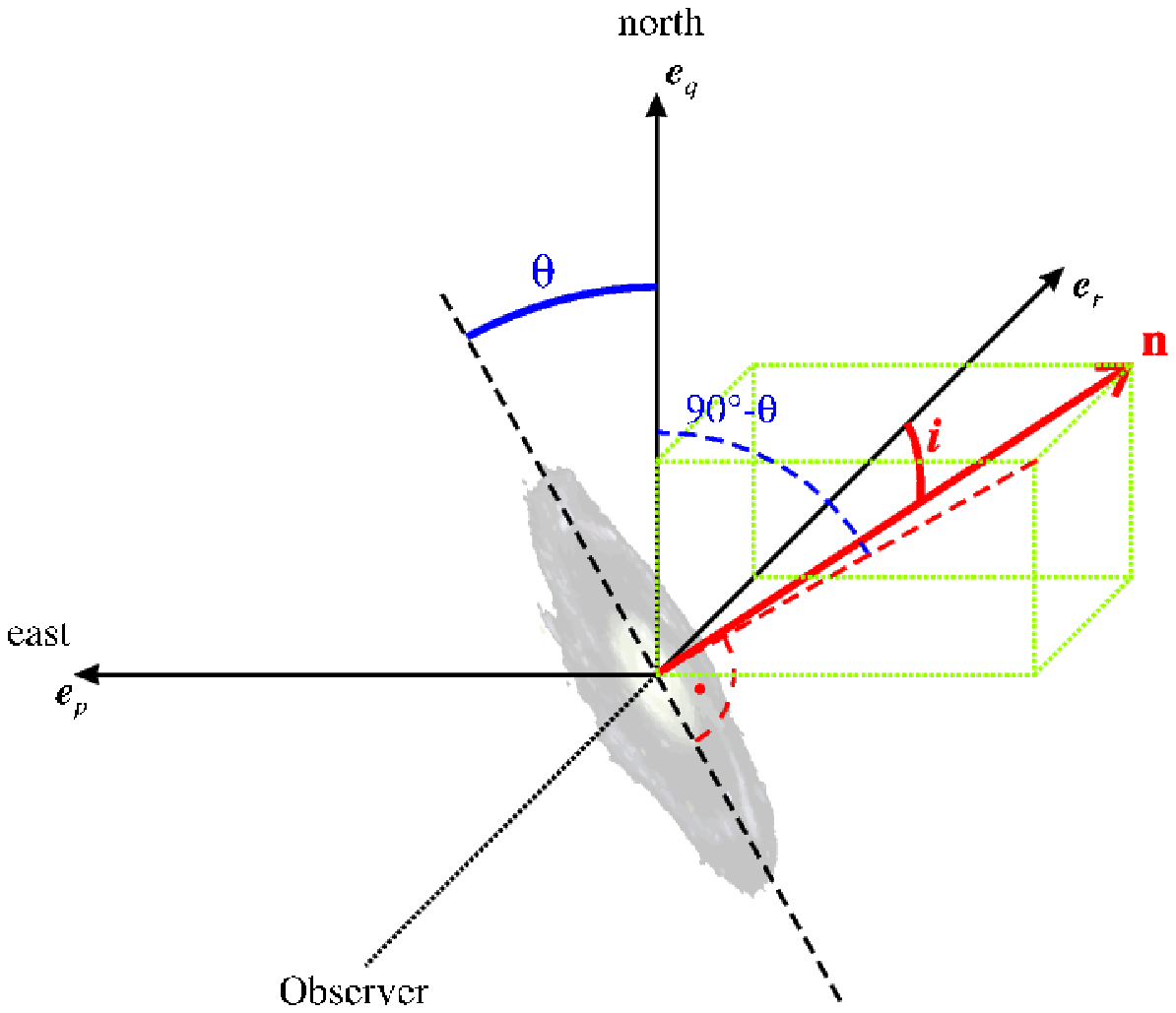} }
  \caption{This artist-view shows the orientation of the normal $\vect{n}$ to the equatorial plane of Andromeda which is used to calculate positions in the Andromeda-centric coordinate system. $\vect{e}_r$ is the direction from the observer to Andromeda. $\vect{e}_p$ is parallel to the celestial equator and pointing towards east and $\vect{e}_q$ is pointing towards the celestial north pole. The inclination $i$ is the angle between the normal of the equatorial plane and the line-of-sight. The position angle $\theta$ is the angle between the projected major axis of the disc of Andromeda and the direction to the north celestial pole, measured from north through east.}
  \label{fig_m31coords}
\end{figure}

To calculate the Cartesian components of a vector in the Andromeda-centric coordinate system, the heliocentric position vector $\vect{r}$ is shifted to the centre of M31 and rotated to the Andromeda-centric coordinate system,
\begin{equation}
\vect{r}_{\rm M31} = \matr{R}_{\rm M31} \left( \vect{r}-\vect{r}(\mbox{\small M31}) \right), \label{eqn_generaltransformation}
\end{equation}
where $\vect{r}(\mbox{\small M31})$ is the heliocentric position vector of M31.

To calculate the rotation matrix $\matr{R}_{\rm M31}$ in Eq.~(\ref{eqn_generaltransformation}) it is not necessary and also not advisable to start with coordinates given in Galactic coordinates $(l,b)$. Rather, it is recommended to work from right ascension and declination ($\alpha$, $\delta$). The reason is the way the orientation of the Andromeda disc is given: the position angle $\theta$ is the angle between a given direction and the direction to the north celestial pole (NCP), conventionally measured from north through east \citep[e.g.,][]{bm98}. Using equatorial coordinates eliminates the need to correct for the difference between the direction to the NCP and the north Galactic pole (NGP) which is in general different for different positions on the celestial sky.

The first step is to calculate the transformation-matrix that rotates the `normal triad' \citep{mu83} parallel to the axes of the Cartesian coordinate system. This is given by the matrix $\matr{R}_{\vect{r}\vect{p}\vect{q}}$, where the columns are the triad of unit-vectors $\vect{e}_r$, $\vect{e}_p$, and $\vect{e}_q$:
\begin{eqnarray}
\matr{R}_{\vect{r}\vect{p}\vect{q}}(\alpha,\delta)
  &=& \left( \vect{e}_r \; \vect{e}_p \; \vect{e}_q \right) \nonumber \\
  &=& \begin{pmatrix}
      \cos\delta\cos\alpha & -\sin\alpha & -\sin\delta\cos\alpha \\
      \cos\delta\sin\alpha &  \cos\alpha & -\sin\delta\sin\alpha \\
      \sin\delta & 0 & \cos\delta
    \end{pmatrix}.
\end{eqnarray}
$\vect{e}_r$ is in general pointing in the direction with equatorial coordinates $(\alpha,\delta)$, $\vect{e}_p$ is parallel to the celestial equator, positive towards the east, and $\vect{e}_q$ is pointing towards the north \citep{mu83}. Here $\left( \vect{e}_r \; \vect{e}_p \; \vect{e}_q \right)$ is the normal triad of the coordinates of M31 $(\alpha_{\rm M31},\delta_{\rm M31})$.

A preliminary rotation matrix is given by:
\begin{equation}
\matr{R}'_{\rm M31} = \matr{R}_y(90\degr-i) \matr{R}_x(90\degr-\theta) \matr{R}_{\vect{r}\vect{p}\vect{q}}(\alpha_{\rm M31},\delta_{\rm M31}) \label{eqn_PTMatrix}
\end{equation}
where $\matr{R}_{x/y/z}(\gamma)$ are the matrices which perform a counter-clockwise rotation about the Cartesian coordinate axes by an angle
$\gamma$, e.g.
\begin{equation}
\matr{R}_x(\gamma) = \begin{pmatrix}
1 & 0 & 0 \\
0 & \cos \gamma & \sin \gamma \\
0 & -\sin \gamma & \cos \gamma
\end{pmatrix}.
\end{equation}
$\matr{R}_x(90\degr-\theta)$ corrects for the position angle and $\matr{R}_y(90\degr-i)$ for the inclination of M31 (Fig.~\ref{fig_m31coords}). This rotation is preliminary because it performs a transformation such that \emph{the Sun} is located at $(l'_{\rm M31},b'_{\rm M31})=(180\degr,-12.5\degr)$ but we defined the Milky Way, i.e.\ the Galactic Centre, to be located at $l_{\rm M31}=0\degr$.
Therefore an additional rotation of the order $\arctan(8.5\unit{kpc}/785\unit{kpc})=0.6\degr$ has to be performed.

The procedure to calculate this angle is as follows: (i)~transform the vector pointing from Andromeda to the GC to the preliminary system
\begin{equation}
\vect{r}'(\mbox{\small GC}) = \matr{R}'_{\rm M31} \left( \vect{r}(\mbox{\small M31}) - 
                                                   \vect{r}(\mbox{\small GC}) \right), \label{eqn_calcbeta_i}
\end{equation}
(ii)~project this vector onto the plane of M31,
\begin{equation}
\vect{r}'_{xy}(\mbox{\small GC})%
  = {\vect{e}'}_{x}^\oAnd ({\vect{e}'}_{x}^\oAnd \cdot \vect{r}'(\mbox{\small GC})) +%
    {\vect{e}'}_{y}^\oAnd ({\vect{e}'}_{y}^\oAnd \cdot \vect{r}'(\mbox{\small GC})), \label{eqn_calcbeta_ii}
\end{equation}
and (iii)~calculate the angle $\beta$ between this vector and the preliminary ${\vect{e}'}_x^\oAnd$-axis:
\begin{equation}
\beta = \arccos \left( %
  \frac{\vect{r}'_{xy}(\mbox{\small GC}) \cdot {\vect{e}'}_{x}^\oAnd}%
       {\left|\vect{r}'_{xy}(\mbox{\small GC}) \right|} \right) \label{eqn_calcbeta_iii}.
\end{equation}
The direction of rotation about the ${\vect{e}'}_{z}^\oAnd$-axis can be calculated by projecting on the ${\vect{e}'}_{y}^\oAnd$-axis:
\begin{equation}
s_z = \frac{
  \vect{r}'_{xy}(\mbox{\small GC}) \cdot {\vect{e}'}_{y}^\oAnd
}{
  | \vect{r}'_{xy}(\mbox{\small GC}) \cdot {\vect{e}'}_{y}^\oAnd | }
= \pm 1.
\end{equation}
Now the full transformation matrix can be calculated:
\begin{eqnarray}
\matr{R}_{\rm M31} &=& \matr{R}_z^{s_{\!z}}(\beta) \matr{R}_y(90\degr-i) \matr{R}_x(90\degr-\theta) \nonumber \\ 
            & & \matr{R}_{\vect{r}\vect{p}\vect{q}}(\alpha_{\rm M31},\delta_{\rm M31})%
            \label{eqn_FTMatrix} \\
&=&
\begin{pmatrix}
0.7703  & 0.3244 & 0.5490 \\
-0.6321 & 0.5017 & 0.5905 \\
-0.0839 & -0.8019 & 0.5915
\end{pmatrix} \label{eqn_RM31matrix}.
\end{eqnarray}
Note that $\matr{R}_z^{-1}(\beta)=\matr{R}_z(-\beta)$. We use an inclination $i=77.5\degr$ and a position angle $\theta=37.7\degr$ \citep{devau58}. To calculate the angle $\beta$ and the matrix $\matr{R}_{\vect{r}\vect{p}\vect{q}}(\alpha_{\rm M31},\delta_{\rm M31})$ in Eq.~(\ref{eqn_RM31matrix}) we use a distance of $785\unit{kpc}$ and coordinates $\alpha_{\rm M31}=00 \hou42 \min 44.3$, $\delta_{\rm M31}=+41 \degr 16 \arcmin 09.4 \arcsec$ for M31, and we employ a distance $8.5\unit{kpc}$ of the Sun from the GC. 

The coordinate transformation as given in \citet{koch06} must be wrong as the orientation of the triad at the position of Andromeda was ignored, just shifting to the centre of M31. The offset of the directions of the NCP and the NGP was also not correctly calculated. This offset is by chance very small at the position of Andromeda on the celestial sky \citep[compare to][]{mccon06}. According to their transformation, Andromeda would have an inclination of $i=66\degr$ and a position angle of $\theta=-16\degr$. However, since they performed only orthogonal transformations, translation and rotation, the relative orientations of the satellites are preserved.

\subsection{General transformation}
Equation (\ref{eqn_generaltransformation}) is the transformation equation performing the necessary linear operations, translation and rotation, respectively. Generally this becomes
\begin{equation}
\vect{r}_{\mathcal{T}} = %
  \matr{R}_{\mathcal{B} \rightarrow \mathcal{T}} %
  \left( \vect{r}_{\mathcal{B}} - \vect{r}_{\mathcal{B}}(\mbox{\small O}_{\mathcal{T}}) \right),
\end{equation}
where $\vect{r}_B$ is the Cartesian vector of any object picked to be transformed to the target system $\mathcal{T}$, $\vect{r}_B(\mbox{\small O}_{\mathcal{T}})$ is the vector of the origin of the target system $\mathcal{T}$, both vectors given in the basis coordinate system $\mathcal{B}$. The appropriate rotation matrix $\matr{R}_{\mathcal{B} \rightarrow \mathcal{T}}$ can be calculated as given in Eq.~(\ref{eqn_PTMatrix}) or (\ref{eqn_FTMatrix}). Equation (\ref{eqn_PTMatrix}) gives the rotation matrix for which the Sun is located at longitude $180\degr$. Equations (\ref{eqn_calcbeta_i}) -- (\ref{eqn_FTMatrix}) show how to derive the rotation matrix such that the GC is at longitude $180\degr$. $\boldsymbol{r}_{\mathcal{T}}$ is the Cartesian vector of the object given in the target coordinate system. Equatorial coordinates follow from Eqs.~(\ref{eqn_lbr1}) -- (\ref{eqn_lbr3}).

\subsection{Example Transformation}\label{sec_apptransfo}
We give here an example of how to transform the coordinates of a companion galaxy into the Andromeda-centric coordinate system as described in Section~\ref{sec_M31coords}. We apply the transformation to M33 with coordinates $\alpha=01 \hou 33 \min 51$, $\delta=30 \degr 39\arcmin 36 \arcsec$, and a distance $D=809\unit{kpc}$ from the Sun, for Andromeda, $\alpha=00 \hou42 \min 44.3$, $\delta=+41 \degr 16 \arcmin 09.4 \arcsec$, with a distance $D=785\unit{kpc}$.  First we need to calculate the Cartesian coordinates of M31 and M33 in the heliocentric coordinate system \citep{mu83}:
\begin{eqnarray}
\vect{r}(\mbox{\small M31}) &=& \begin{pmatrix}579.791 & 109.391 & 517.785\end{pmatrix}\transpose \; , \nonumber\\
\vect{r}(\mbox{\small M33}) &=& \begin{pmatrix}638.372 & 277.075 & 412.543\end{pmatrix}\transpose \; , \nonumber
\end{eqnarray}
where $\transpose$ denotes transposition.
Next we apply Eq.~(\ref{eqn_generaltransformation}) using the matrix $\matr{R}_{\rm M31}$ as given in Eq.~(\ref{eqn_RM31matrix}):
\begin{eqnarray}
\vect{r}_{\rm M31} &=& \matr{R}_{\rm M31}( \vect{r}(\mbox{\small M33}) - \vect{r}(\mbox{\small M31}) ) \; , \nonumber \\
 &=& \begin{pmatrix} 41.737 & -15.051 & -201.635 \end{pmatrix}\transpose \; , \nonumber
\end{eqnarray}
and finally we can calculate the galactic coordinates in the Andromeda-centric coordinate system (Eqs.~\ref{eqn_lbr1}--\ref{eqn_lbr3}): 
\begin{eqnarray}
l_{\rm M31} &=& 340.2\degr\;, \nonumber \\
b_{\rm M31} &=& -77.6\degr\;, \nonumber \\
r_{\rm M31} &=& 206.5\unit{kpc} \; \nonumber.
\end{eqnarray}

\section{Software implementation}
All algorithms are implemented using the Python programming language (\url{http://www.python.org}) and extending it using C routines. We use the packages Numeric (\url{http://numpy.scipy.org}) and its implementations of basic linear algebra routines as well as the SciPy package (\url{http://www.scipy.org}). The ODR method (\url{http://www.netlib.org/odrpack}) is implemented using the wrapper-package odr by Robert Kern (\url{http://www.python.net/crew/kernr/}).

The software is available at \url{http://www.astro.uni-bonn.de/downloads} released under the GNU General Public Licence (GPL).

\bibliographystyle{mn2e}
\bibliography{}

\label{lastpage}
\end{document}

%% file: mkj.bbl
\begin{thebibliography}{}

\bibitem[\protect\citeauthoryear{{Bekki}, {Couch}, {Drinkwater} \&
  {Gregg}}{{Bekki} et~al.}{2001}]{bekki01}
{Bekki} K.,  {Couch} W.~J.,  {Drinkwater} M.~J.,    {Gregg} M.~D.,  2001,
  \apjl, 557, L39

\bibitem[\protect\citeauthoryear{Belokurov, Zucker, Evans, Kleyna, Koposov
  et~al.,}{Belokurov et~al.}{2006}]{belok06a}
Belokurov V.,  Zucker D.~B.,  Evans N.~W.,  Kleyna J.~T.,  Koposov S.,
  et~al., 2006, preprint astro-ph/0608448

\bibitem[\protect\citeauthoryear{{Belokurov}, {Zucker}, {Evans}, {Wilkinson},
  {Irwin} et~al.,}{{Belokurov} et~al.}{2006}]{belok06}
{Belokurov} V.,  {Zucker} D.~B.,  {Evans} N.~W.,  {Wilkinson} M.~I.,  {Irwin}
  M.~J.,    et~al., 2006, \apjl, 647, L111

\bibitem[\protect\citeauthoryear{Binney \& Merryfield}{Binney \&
  Merryfield}{1998}]{bm98}
Binney J.,  Merryfield M.,  1998, Galactic Astronomy.
Princeton University Press

\bibitem[\protect\citeauthoryear{{Brunthaler}, {Reid}, {Falcke}, {Greenhill} \&
  {Henkel}}{{Brunthaler} et~al.}{2005}]{brunt05}
{Brunthaler} A.,  {Reid} M.~J.,  {Falcke} H.,  {Greenhill} L.~J.,    {Henkel}
  C.,  2005, Science, 307, 1440

\bibitem[\protect\citeauthoryear{{Bullock}}{{Bullock}}{2002}]{bullo02}
{Bullock} J.~S.,  2002, in {Natarajan} P.,  ed., The shapes of galaxies and
  their dark halos, Proceedings of the Yale Cosmology Workshop ''The Shapes of
  Galaxies and Their Dark Matter Halos'', New Haven, Connecticut, USA, 28-30
  May 2001. Edited by Priyamvada Natarajan. Singapore: World Scientific, 2002,
  ISBN 9810248482, p.109 {Shapes of dark matter halos}.
pp 109--+

\bibitem[\protect\citeauthoryear{Chojnacki, Brooks, {van den Hengel} \&
  Gawley}{Chojnacki et~al.}{2000}]{chojn00}
Chojnacki W.,  Brooks M.,  {van den Hengel} A.,    Gawley D.,  2000, {IEEE
  Transactions on pattern analysis and machine intelligence}, 22, 1294

\bibitem[\protect\citeauthoryear{{de Vaucouleurs}}{{de
  Vaucouleurs}}{1958}]{devau58}
{de Vaucouleurs} G.,  1958, \apj, 128, 465

\bibitem[\protect\citeauthoryear{{de Vaucouleurs}, {de Vaucouleurs}, {Corwin}
  Jr., {Buta}, {Paturel} \& {Fouque}}{{de Vaucouleurs} et~al.}{1991}]{devau91}
{de Vaucouleurs} G.,  {de Vaucouleurs} A.,  {Corwin} Jr. H.~G.,  {Buta} R.~J.,
  {Paturel} G.,    {Fouque} P.,  1991, {Third Reference Catalogue of Bright
  Galaxies}.
Volume 1-3, XII, 2069 pp.~7 figs..~ Springer-Verlag Berlin Heidelberg New York

\bibitem[\protect\citeauthoryear{{Doroshkevich}}{{Doroshkevich}}{1973}]{doros7%
3}
{Doroshkevich} A.~G.,  1973, \aplett, 14, 11

\bibitem[\protect\citeauthoryear{{Doroshkevich}, {Shandarin} \&
  {Saar}}{{Doroshkevich} et~al.}{1978}]{doros78}
{Doroshkevich} A.~G.,  {Shandarin} S.~F.,    {Saar} E.,  1978, \mnras, 184, 643

\bibitem[\protect\citeauthoryear{{Dubinski} \& {Carlberg}}{{Dubinski} \&
  {Carlberg}}{1991}]{dubin91}
{Dubinski} J.,  {Carlberg} R.~G.,  1991, \apj, 378, 496

\bibitem[\protect\citeauthoryear{{Einasto} \& {Lynden-Bell}}{{Einasto} \&
  {Lynden-Bell}}{1982}]{einas82}
{Einasto} J.,  {Lynden-Bell} D.,  1982, \mnras, 199, 67

\bibitem[\protect\citeauthoryear{Fellhauer, Belokurov, Evans, Wilkinson, Zucker
  et~al.,}{Fellhauer et~al.}{2006}]{fellh06}
Fellhauer M.,  Belokurov V.,  Evans N.~W.,  Wilkinson M.~I.,  Zucker D.~B.,
  et~al., 2006, preprint astro-ph/0605026

\bibitem[\protect\citeauthoryear{Fisher, Lewis \& Embleton}{Fisher
  et~al.}{1987}]{fi87}
Fisher N.,  Lewis T.,    Embleton B.,  1987, Statistical analysis of spherical
  data.
{Cambridge University Press}

\bibitem[\protect\citeauthoryear{{Franx}, {Illingworth} \& {de Zeeuw}}{{Franx}
  et~al.}{1991}]{franx91}
{Franx} M.,  {Illingworth} G.,    {de Zeeuw} T.,  1991, \apj, 383, 112

\bibitem[\protect\citeauthoryear{Gilmore, Wilkinson, Kleyna, Koch, Evans
  et~al.,}{Gilmore et~al.}{2006}]{gilmo06}
Gilmore G.,  Wilkinson M.,  Kleyna J.,  Koch A.,  Evans N.~W.,    et~al., 2006,
  in {UCLA Dark Matter 2006 conference} {Observed Properties of Dark Matter:
  dynamical studies of dSph galaxies}

\bibitem[\protect\citeauthoryear{{Gnedin} \& {Zhao}}{{Gnedin} \&
  {Zhao}}{2002}]{gnedi02}
{Gnedin} O.~Y.,  {Zhao} H.,  2002, \mnras, 333, 299

\bibitem[\protect\citeauthoryear{{Governato}, {Mayer}, {Wadsley}, {Gardner},
  {Willman}, {Hayashi}, {Quinn}, {Stadel} \& {Lake}}{{Governato}
  et~al.}{2004}]{gover04}
{Governato} F.,  {Mayer} L.,  {Wadsley} J.,  {Gardner} J.~P.,  {Willman} B.,
  {Hayashi} E.,  {Quinn} T.,  {Stadel} J.,    {Lake} G.,  2004, \apj, 607, 688

\bibitem[\protect\citeauthoryear{{Grebel}, {Kolatt} \& {Brandner}}{{Grebel}
  et~al.}{1999}]{grebe99}
{Grebel} E.~K.,  {Kolatt} T.,    {Brandner} W.,  1999, in {Whitelock} P.,
  {Cannon} R.,  eds, IAU Symposium {Orbits versus Star Formation Histories: A
  Progress Report}.
pp 447--+

\bibitem[\protect\citeauthoryear{{Hartwick}}{{Hartwick}}{2000}]{hartw00}
{Hartwick} F.~D.~A.,  2000, \aj, 119, 2248

\bibitem[\protect\citeauthoryear{{Herrmann} \& {Ciardullo}}{{Herrmann} \&
  {Ciardullo}}{2005}]{herrm05}
{Herrmann} K.~A.,  {Ciardullo} R.,  2005, American Astronomical Society Meeting
  Abstracts, 207,

\bibitem[\protect\citeauthoryear{{Hu}, {Yuan}, {Su}, {Wu} \& {Liu}}{{Hu}
  et~al.}{1998}]{hu98}
{Hu} F.~X.,  {Yuan} Q.~R.,  {Su} H.~J.,  {Wu} G.~X.,    {Liu} Y.~Z.,  1998,
  \apj, 495, 179

\bibitem[\protect\citeauthoryear{{Hunsberger}, {Charlton} \&
  {Zaritsky}}{{Hunsberger} et~al.}{1996}]{hunsb96}
{Hunsberger} S.~D.,  {Charlton} J.~C.,    {Zaritsky} D.,  1996, \apj, 462, 50

\bibitem[\protect\citeauthoryear{{Hunter}, {Hunsberger} \& {Roye}}{{Hunter}
  et~al.}{2000}]{hunte00}
{Hunter} D.~A.,  {Hunsberger} S.~D.,    {Roye} E.~W.,  2000, \apj, 542, 137

\bibitem[\protect\citeauthoryear{{Jing} \& {Suto}}{{Jing} \&
  {Suto}}{2002}]{jing02}
{Jing} Y.~P.,  {Suto} Y.,  2002, \apj, 574, 538

\bibitem[\protect\citeauthoryear{{Kahn} \& {Woltjer}}{{Kahn} \&
  {Woltjer}}{1959}]{kahn59}
{Kahn} F.~D.,  {Woltjer} L.,  1959, \apj, 130, 705

\bibitem[\protect\citeauthoryear{Kang, Mao, Gao \& Jing}{Kang
  et~al.}{2005}]{kang05}
Kang X.,  Mao S.,  Gao L.,    Jing Y.,  2005, \aap, 437, 383

\bibitem[\protect\citeauthoryear{{Kase}, {Makino} \& {Funato}}{{Kase}
  et~al.}{2006}]{kase06}
{Kase} H.,  {Makino} J.,    {Funato} Y.,  2006, in press astro-ph/0603074

\bibitem[\protect\citeauthoryear{{Kazantzidis}, {Mayer}, {Mastropietro},
  {Diemand}, {Stadel} \& {Moore}}{{Kazantzidis} et~al.}{2004}]{kazan04a}
{Kazantzidis} S.,  {Mayer} L.,  {Mastropietro} C.,  {Diemand} J.,  {Stadel} J.,
     {Moore} B.,  2004, \apj, 608, 663

\bibitem[\protect\citeauthoryear{{Kleyna}, {Wilkinson}, {Gilmore} \&
  {Evans}}{{Kleyna} et~al.}{2003}]{kleyn03}
{Kleyna} J.~T.,  {Wilkinson} M.~I.,  {Gilmore} G.,    {Evans} N.~W.,  2003,
  \apjl, 588, L21

\bibitem[\protect\citeauthoryear{{Klypin}, {Kravtsov}, {Valenzuela} \&
  {Prada}}{{Klypin} et~al.}{1999}]{klypi99}
{Klypin} A.,  {Kravtsov} A.~V.,  {Valenzuela} O.,    {Prada} F.,  1999, \apj,
  522, 82

\bibitem[\protect\citeauthoryear{{Koch} \& {Grebel}}{{Koch} \&
  {Grebel}}{2006}]{koch06}
{Koch} A.,  {Grebel} E.~K.,  2006, \aj, 131, 1405

\bibitem[\protect\citeauthoryear{{Kroupa}}{{Kroupa}}{1998}]{kroup98b}
{Kroupa} P.,  1998, in {The Magellanic Clouds and Other Dwarf Galaxies} {dSph
  Satellite Galaxies without Dark Matter: a Study of Parameter Space}

\bibitem[\protect\citeauthoryear{{Kroupa}, {Theis} \& {Boily}}{{Kroupa}
  et~al.}{2005}]{kroup05}
{Kroupa} P.,  {Theis} C.,    {Boily} C.~M.,  2005, \aap, 431, 517

\bibitem[\protect\citeauthoryear{{Kunkel} \& {Demers}}{{Kunkel} \&
  {Demers}}{1976}]{kunke76}
{Kunkel} W.~E.,  {Demers} S.,  1976, in The Galaxy and the Local Group {The
  Magellanic Plane}.
pp 241--+

\bibitem[\protect\citeauthoryear{{Lee} \& {Kim}}{{Lee} \& {Kim}}{2000}]{lee00}
{Lee} M.~G.,  {Kim} S.~C.,  2000, \aj, 119, 777

\bibitem[\protect\citeauthoryear{{Libeskind}, {Frenk}, {Cole}, {Helly},
  {Jenkins}, {Navarro} \& {Power}}{{Libeskind} et~al.}{2005}]{libes05}
{Libeskind} N.~I.,  {Frenk} C.~S.,  {Cole} S.,  {Helly} J.~C.,  {Jenkins} A.,
  {Navarro} J.~F.,    {Power} C.,  2005, \mnras, pp 783--+

\bibitem[\protect\citeauthoryear{{Loeb}, {Reid}, {Brunthaler} \&
  {Falcke}}{{Loeb} et~al.}{2005}]{loeb05}
{Loeb} A.,  {Reid} M.~J.,  {Brunthaler} A.,    {Falcke} H.,  2005, \apj, 633,
  894

\bibitem[\protect\citeauthoryear{{Lynden-Bell}}{{Lynden-Bell}}{1976}]{lynde76}
{Lynden-Bell} D.,  1976, \mnras, 174, 695

\bibitem[\protect\citeauthoryear{{Lynden-Bell} \& {Lynden-Bell}}{{Lynden-Bell}
  \& {Lynden-Bell}}{1995}]{lynde95}
{Lynden-Bell} D.,  {Lynden-Bell} R.~M.,  1995, \mnras, 275, 429

\bibitem[\protect\citeauthoryear{{Majewski}}{{Majewski}}{1994}]{majew94}
{Majewski} S.~R.,  1994, \apjl, 431, L17

\bibitem[\protect\citeauthoryear{{Martin}, {Ibata}, {Bellazzini}, {Irwin},
  {Lewis} \& {Dehnen}}{{Martin} et~al.}{2004}]{marti04}
{Martin} N.~F.,  {Ibata} R.~A.,  {Bellazzini} M.,  {Irwin} M.~J.,  {Lewis}
  G.~F.,    {Dehnen} W.,  2004, \mnras, 348, 12

\bibitem[\protect\citeauthoryear{{Martin}, {Ibata}, {Irwin}, {Chapman},
  {Lewis}, {Ferguson}, {Tanvir} \& {McConnachie}}{{Martin}
  et~al.}{2006}]{marti06}
{Martin} N.~F.,  {Ibata} R.~A.,  {Irwin} M.~J.,  {Chapman} S.,  {Lewis} G.~F.,
  {Ferguson} A.~M.~N.,  {Tanvir} N.,    {McConnachie} A.~W.,  2006, \mnras, pp
  947--+

\bibitem[\protect\citeauthoryear{{Mateo}}{{Mateo}}{1998}]{mateo98}
{Mateo} M.~L.,  1998, \araa, 36, 435

\bibitem[\protect\citeauthoryear{{McConnachie} \& {Irwin}}{{McConnachie} \&
  {Irwin}}{2006a}]{mccon06a}
{McConnachie} A.~W.,  {Irwin} M.~J.,  2006a, \mnras, 365, 1263

\bibitem[\protect\citeauthoryear{{McConnachie} \& {Irwin}}{{McConnachie} \&
  {Irwin}}{2006b}]{mccon06}
{McConnachie} A.~W.,  {Irwin} M.~J.,  2006b, \mnras, 365, 902

\bibitem[\protect\citeauthoryear{{McConnachie}, {Irwin}, {Ferguson}, {Ibata},
  {Lewis} \& {Tanvir}}{{McConnachie} et~al.}{2005}]{mccon05}
{McConnachie} A.~W.,  {Irwin} M.~J.,  {Ferguson} A.~M.~N.,  {Ibata} R.~A.,
  {Lewis} G.~F.,    {Tanvir} N.,  2005, \mnras, 356, 979

\bibitem[\protect\citeauthoryear{{Moitinho}, {V{\'a}zquez}, {Carraro}, {Baume},
  {Giorgi} \& {Lyra}}{{Moitinho} et~al.}{2006}]{moiti06}
{Moitinho} A.,  {V{\'a}zquez} R.~A.,  {Carraro} G.,  {Baume} G.,  {Giorgi}
  E.~E.,    {Lyra} W.,  2006, \mnras, 368, L77

\bibitem[\protect\citeauthoryear{{Moore}, {Ghigna}, {Governato}, {Lake},
  {Quinn}, {Stadel} \& {Tozzi}}{{Moore} et~al.}{1999}]{moore99}
{Moore} B.,  {Ghigna} S.,  {Governato} F.,  {Lake} G.,  {Quinn} T.,  {Stadel}
  J.,    {Tozzi} P.,  1999, \apjl, 524, L19

\bibitem[\protect\citeauthoryear{{Mu{\~n}oz}, {Frinchaboy}, {Majewski}, {Kuhn},
  {Chou}, {Palma}, {Sohn}, {Patterson} \& {Siegel}}{{Mu{\~n}oz}
  et~al.}{2005}]{munoz05}
{Mu{\~n}oz} R.~R.,  {Frinchaboy} P.~M.,  {Majewski} S.~R.,  {Kuhn} J.~R.,
  {Chou} M.-Y.,  {Palma} C.,  {Sohn} S.~T.,  {Patterson} R.~J.,    {Siegel}
  M.~H.,  2005, \apjl, 631, L137

\bibitem[\protect\citeauthoryear{{Murray}}{{Murray}}{1983}]{mu83}
{Murray} C.~A.,  1983, {Vectorial astrometry}.
Bristol: Adam Hilger, 1983

\bibitem[\protect\citeauthoryear{{Navarro}, {Abadi} \& {Steinmetz}}{{Navarro}
  et~al.}{2004}]{navar04}
{Navarro} J.~F.,  {Abadi} M.~G.,    {Steinmetz} M.,  2004, \apjl, 613, L41

\bibitem[\protect\citeauthoryear{{Newberg}, {Yanny}, {Rockosi}, {Grebel}, {Rix}
  et~al.,}{{Newberg} et~al.}{2002}]{newbe02}
{Newberg} H.~J.,  {Yanny} B.,  {Rockosi} C.,  {Grebel} E.~K.,  {Rix}   et~al.,
  2002, \apj, 569, 245

\bibitem[\protect\citeauthoryear{{Palma}, {Majewski} \& {Johnston}}{{Palma}
  et~al.}{2002}]{palma02}
{Palma} C.,  {Majewski} S.~R.,    {Johnston} K.~V.,  2002, \apj, 564, 736

\bibitem[\protect\citeauthoryear{{Pe{\~ n}arrubia}, {Kroupa} \& {Boily}}{{Pe{\~
  n}arrubia} et~al.}{2002}]{penar02}
{Pe{\~ n}arrubia} J.,  {Kroupa} P.,    {Boily} C.~M.,  2002, \mnras, 333, 779

\bibitem[\protect\citeauthoryear{{Read}, {Pontzen} \& {Viel}}{{Read}
  et~al.}{2006}]{read06}
{Read} J.~I.,  {Pontzen} A.~P.,    {Viel} M.,  2006, \mnras, 371, 885

\bibitem[\protect\citeauthoryear{Recchi, Kroupa, Theis \& Hensler}{Recchi
  et~al.}{2006}]{recch06a}
Recchi S.,  Kroupa P.,  Theis C.,    Hensler G.,  2006, submitted

\bibitem[\protect\citeauthoryear{{Schmidt}, {Keller}, {Francis} \&
  {Bessell}}{{Schmidt} et~al.}{2005}]{schmi05}
{Schmidt} B.~P.,  {Keller} S.~C.,  {Francis} P.~J.,    {Bessell} M.~S.,  2005,
  American Astronomical Society Meeting Abstracts, 206,

\bibitem[\protect\citeauthoryear{{Stoehr}, {White}, {Tormen} \&
  {Springel}}{{Stoehr} et~al.}{2002}]{stoeh02}
{Stoehr} F.,  {White} S.~D.~M.,  {Tormen} G.,    {Springel} V.,  2002, \mnras,
  335, L84

\bibitem[\protect\citeauthoryear{{van den Bergh}}{{van den
  Bergh}}{1999}]{vdber99}
{van den Bergh} S.,  1999, \aapr, 9, 273

\bibitem[\protect\citeauthoryear{Walter, Martin \& Ott}{Walter
  et~al.}{2006}]{walte06}
Walter F.,  Martin C.~L.,    Ott J.,  2006, preprint astro-ph/0608169

\bibitem[\protect\citeauthoryear{{Weilbacher}, {Duc} \&
  {Fritze-v.~Alvensleben}}{{Weilbacher} et~al.}{2003}]{weilb03}
{Weilbacher} P.~M.,  {Duc} P.-A.,    {Fritze-v.~Alvensleben} U.,  2003, \aap,
  397, 545

\bibitem[\protect\citeauthoryear{{Wilkinson}, {Kleyna}, {Evans} \&
  {Gilmore}}{{Wilkinson} et~al.}{2002}]{wilki02}
{Wilkinson} M.~I.,  {Kleyna} J.,  {Evans} N.~W.,    {Gilmore} G.,  2002,
  \mnras, 330, 778

\bibitem[\protect\citeauthoryear{{Wilkinson}, {Kleyna}, {Wyn Evans}, {Gilmore},
  {Read} et~al.,}{{Wilkinson} et~al.}{2006}]{wilki06}
{Wilkinson} M.~I.,  {Kleyna} J.~T.,  {Wyn Evans} N.,  {Gilmore} G.~F.,  {Read}
  J.~I.,    et~al., 2006, in EAS Publications Series {The internal kinematics
  of dwarf spheroidal galaxies}.
pp 105--112

\bibitem[\protect\citeauthoryear{{Willman}, {Blanton}, {West}, {Dalcanton},
  {Hogg} et~al.,}{{Willman} et~al.}{2005}]{willm05a}
{Willman} B.,  {Blanton} M.~R.,  {West} A.~A.,  {Dalcanton} J.~J.,  {Hogg}
  D.~W.,    et~al., 2005, \aj, 129, 2692

\bibitem[\protect\citeauthoryear{{Willman}, {Dalcanton}, {Martinez-Delgado},
  {West}, {Blanton} et~al.,}{{Willman} et~al.}{2005}]{willm05}
{Willman} B.,  {Dalcanton} J.~J.,  {Martinez-Delgado} D.,  {West} A.~A.,
  {Blanton} M.~R.,    et~al., 2005, \apjl, 626, L85

\bibitem[\protect\citeauthoryear{{York}, {Adelman}, {Anderson} \& {et
  al.}}{{York} et~al.}{2000}]{york00}
{York} D.~G.,  {Adelman} J.,  {Anderson} J.~E.,    {et al.} 2000, \aj, 120,
  1579

\bibitem[\protect\citeauthoryear{{Zentner}, {Kravtsov}, {Gnedin} \&
  {Klypin}}{{Zentner} et~al.}{2005}]{zentn05}
{Zentner} A.~R.,  {Kravtsov} A.~V.,  {Gnedin} O.~Y.,    {Klypin} A.~A.,  2005,
  \apj, 629, 219

\bibitem[\protect\citeauthoryear{{Zucker}, {Belokurov}, {Evans}, {Wilkinson},
  {Irwin} et~al.,}{{Zucker} et~al.}{2006}]{zucke06}
{Zucker} D.~B.,  {Belokurov} V.,  {Evans} N.~W.,  {Wilkinson} M.~I.,  {Irwin}
  et~al., 2006, \apjl, 643, L103

\bibitem[\protect\citeauthoryear{{Zucker}, {Kniazev}, {Bell},
  {Mart{\'{\i}}nez-Delgado}, {Grebel} \& {et~al.}}{{Zucker}
  et~al.}{2004}]{zucke04}
{Zucker} D.~B.,  {Kniazev} A.~Y.,  {Bell} E.~F.,  {Mart{\'{\i}}nez-Delgado} D.,
   {Grebel} E.~K.,    {et~al.} 2004, \apjl, 612, L121

\bibitem[\protect\citeauthoryear{Zucker, Kniazev, Mart{\'{\i}}nez-Delgado,
  Bell, Rix et~al.,}{Zucker et~al.}{2006}]{zucke06b}
Zucker D.~B.,  Kniazev A.~Y.,  Mart{\'{\i}}nez-Delgado D.,  Bell E.~F.,  Rix
  H.-W.,    et~al., 2006, preprint astro-ph/0601599

\end{thebibliography}
